\documentclass[12pt]{article} 
\usepackage{times} 
\usepackage{amssymb,epsf}

\newcommand{\bC}{{\mathbb C}}

\newcommand{\bN}{{\mathbb N}}

\newcommand{\bP}{{\mathbb P}}

\newcommand{\bR}{{\mathbb R}}

\newcommand{\bZ}{{\mathbb Z}}
\newcommand{\al}{\alpha}
\newcommand{\be}{\beta}

\newcommand{\de}{\delta}

\newcommand{\ze}{\zeta}
\newcommand{\la}{\lambda}

\newcommand{\si}{\sigma}
\newcommand{\ta}{\tau}

\newcommand{\ch}{\chi}

\newcommand{\om}{\omega}
\newcommand{\vphi}{\varphi}

\newcommand{\veps}{\varepsilon}
\newcommand{\Ga}{\Gamma}
\newcommand{\De}{\Delta}

\newcommand{\Th}{\Theta}

\newcommand{\Si}{\Sigma}

\newcommand{\Ps}{\Psi}

\newcommand{\bGa}{{{\rm I}\kern-.16em \Gamma}}
\newcommand{\cD}{{\cal D}}

\newcommand{\cF}{{\cal F}}
\newcommand{\cG}{{\cal G}}

\newcommand{\cI}{{\cal I}}

\newcommand{\cK}{{\cal K}}
\newcommand{\cL}{{\cal L}}

\newcommand{\cN}{{\cal N}}

\newcommand{\cP}{{\cal P}}
\newcommand{\cQ}{{\cal Q}}

\newcommand{\cT}{{\cal T}}

\newcommand{\cV}{{\cal V}}
\newcommand{\cW}{{\cal W}}
\newcommand{\cX}{{\cal X}}

\newcommand{\db}{{\mkern2mu\mathchar'26\mkern-2mu\mkern-9mud}}

\newcommand{\del}{\partial}

\newcommand{\abs}[1]{{\left\vert #1 \right\vert}}
\newcommand{\norm}[1]{{\left\Vert #1 \right\Vert}}

\newcommand{\Ref}[1]{$(\ref{#1})$}
\newcommand{\sfrac}[2]{{\textstyle \frac{#1}{#2}}}

\newcommand{\const}{\hbox{ \rm const }}

\newcommand{\half}{\frac{1}{2}}

\newcommand{\sli}{\sum\limits}
\newcommand{\pli}{\prod\limits}

\newif\ifintrmk
\def\suffix{ps}
\newcount\system
\global\system=3   % for unix(dvips)
\def\ifundefined#1{\expandafter\ifx\csname#1\endcsname\relax}
\ifundefined{figdir}\def\figdir{}\fi
\newcount\firstline
\newdimen\pswidth  \newdimen\xleft
\newdimen\psheight \newdimen\ytop \newdimen\ybot
\newcount\justx \newcount\justy
\global\justx=0 \global\justy=0
\newdimen\vpos \newtoks\labeL 
\newread\labeLfile \newdimen\xcoord \newdimen\ycoord
\newif\ifdoit 
\newbox\labox
\newdimen\xdvikwid 
\newdimen\xdvikht
\newdimen\pspoints
\newdimen\rwi
\newcount\temp
\def\readdim#1{\global\read\labeLfile to \temp
\global #1=\temp pt}
\def\figcrop#1{\par%  #1=filename
\openin\labeLfile=\figdir#1.lbl                                              
\global\read\labeLfile to\firstline\message{#1}               
\global\read\labeLfile to\temp%read overall dimensions                                     
\readdim{\ybot}
\readdim{\xleft}%               read upper left point
\readdim{\ytop}
\global\read\labeLfile to\justx%ignore
\global\read\labeLfile to\justy%ignore
\global\read\labeLfile to\labeL%ignore
\readdim{\pswidth}%            read lower right point
\global\advance\pswidth by -\xleft
\readdim{\psheight}
\global\advance\ybot by -\psheight
\global\advance\psheight by -\ytop
\global\read\labeLfile to\justx%ignore
\global\read\labeLfile to\justy%ignore
\global\read\labeLfile to\labeL%ignore                                    
\vbox to\psheight{\vfill
\ifnum\system=1
\fi %textures
\ifnum\system=2
\fi %msdos
\ifnum\system=3
                                                 \fi         %%unix:dvips
\ifnum\system=4
\fi         %%unix:dvips,scaled
\ifnum\system=1
\hbox to \pswidth{\kern-\xleft\special{postscriptfile \figdir#1.\suffix }\hfil}\fi
\ifnum\system=2
\hbox to \pswidth{\kern-\xleft\special{ps: plotfile \figdir#1.\suffix }\hfil}\fi
\ifnum\system=3
\hbox to \pswidth{\kern-\xleft\includegraphics{\figdir#1.\suffix}\hfil}\fi
\ifnum\system=4
\hbox to \pswidth{\kern-\xleft\includegraphics{\figdir#1.\suffix}\hfil}\fi
\ifnum\system=5
\hbox to \pswidth{\kern-\xleft\includegraphics{\figdir#1.\suffix}\hfil}\fi %orphee
\ifnum\system=6
   \xdvikwid=\pswidth
   \xdvikht=\psheight
   {\global\divide\xdvikwid by \pspoints}
   {\global\divide\xdvikht by \pspoints}
   \rwi=\xdvikwid
    {\global\multiply\rwi by 10}
\hbox to \pswidth{\kern-\xleft\includegraphics{\figdir#1.\suffix\space}\hfil}\fi                   %xdvik
\vskip -\baselineskip
\vskip -\ybot 
\vskip-\psheight %                                     
\hbox to\pswidth  {\hss}%                                            
\parindent=0pt\offinterlineskip                                       
\vpos=0 pt%                                                              
\loop\readdim{\xcoord}                                 
\ifdim \xcoord < -999pt \doitfalse\else\doittrue\fi                        
\ifdoit \advance \xcoord by -\xleft
\readdim{\ycoord}
\advance \ycoord by -\ytop                              
\global\read\labeLfile to\justx                                       
\global\read\labeLfile to\justy                                       
\global\read\labeLfile to\labeL
\global\setbox\labox=\hbox{\labeL\hskip-0.3em}%    
\advance\vpos by-\ycoord                                              
\vskip-\vpos \vpos=\ycoord                                         
\hbox to\pswidth{\hskip\xcoord %                                 
\hbox to 0pt{\ifnum\justx>0\hss\fi%                                   
\vbox to0pt{%                                                         
\ifnum\justy<2\vss\fi%                                                
\copy\labox\kern0pt%  
\ifnum\justy>0\vss\fi}%                                               
\ifnum\justx<2\hss\fi}%                                               
\hss}%                                                                
\repeat%                                                              
\advance\vpos by-\psheight%                                           
\vskip-\vpos %                                                     
}\closein\labeLfile}
\def\figplace#1#2#3{
\openin\labeLfile=\figdir#1.lbl
\ifeof \labeLfile
       \immediate\write16{***Can't find \figdir#1.lbl; Skipping it.***}
\else  \closein\labeLfile
       \null\hskip#2\raise #3 \hbox{\figcrop{#1}}
\fi
}
\begin{document}
\intrmkfalse
\newcommand{\E}{{\rm e}}
\newcommand{\I}{{\rm i}}
\newcommand{\dd}{{\rm d}}
\newcommand{\az}{a_0}
\newcommand{\aw}{a}
\newcommand{\laz}{\la_0}
\newcommand{\dill}{b}
\newcommand{\tP}{\tilde P}
\newcommand{\Lap}{\De}
\newcommand{\delb}{\bar\del}

\newcommand{\Geff}{\cG_{\rm eff}}
\newcommand{\ETA}{\eta}
\newcommand{\gchi}{\chi}
\newcommand{\lchi}{\tilde\chi}
\newcommand{\Escale}{\epsilon} %\Lambda}
\newcommand{\Esc}[1]{\epsilon_{#1}}
\newcommand{\dpo}[1]{{\rm #1}}
\newcommand{\Tad}{{\cL}}

\newcommand{\Heins}{H1} %{$\bH 1$} %{{\bf H1}}
\newcommand{\Hzwei}{H2} %{$\bH 2$} %{{\bf H2}}
\newcommand{\Hdrei}{H3} %{$\bH 3$} %{{\bf H3}}
\newcommand{\Tr}{\mbox{ Tr}\;}

\newcommand{\Ebrackl}{{\lbrack\!\lbrack}}
\newcommand{\Ebrackr}{{\rbrack\!\rbrack}}
\newcommand{\Equal}[1]{\Ebrackl #1 \Ebrackr}
\newcommand{\BEqual}[1]{{\Big\lbrack}\!\!{\Big\lbrack} #1 {\Big\rbrack}\!\!{\Big\rbrack}}

\newcommand{\Leval}{\Ebrackl}
\newcommand{\Reval}{\Ebrackr}

\newcommand{\Four}{\cI^{(\ge 4)}}

%
%     environments
%
\newtheorem{satz}{Theorem}[section]
\newtheorem{definition}[satz]{Definition}
\newtheorem{lemma}[satz]{Lemma}
\newtheorem{koro}[satz]{Corollary}
\newtheorem{bemerkung}[satz]{Remark}
%
% remove the following if using amsart class
%
\newenvironment{proof}{\par\noindent {\it Proof:} \hspace{7pt}}%
{\hfill\hbox{\vrule width 7pt depth 0pt height 7pt} \par\vspace{10pt}}

%
% put in the following if using amsart
%
%\renewcommand{\cD}{\mathcal{D}}
%\renewcommand{\cL}{\mathcal{L}}

%
%
% Here comes the text
%
%
\title{Dynamical Adjustment of Propagators \\ in Renormalization Group Flows}
\author{Manfred Salmhofer\footnote{salmhofer@itp.uni-leipzig.de}
\footnote{Supported in part by DFG grant Sa 1362/1--1, NSERC and an ESI senior fellowship}
\\  \small 
Institut f\" ur Theoretische Physik, Universit\" at Leipzig,  
\\[-0.5ex] \small
Postfach 100920, 04009 Leipzig, Germany 
\\
\small 
and
\\
\small 
Mathematics Department, University of British Columbia\\
\small
1984 Mathematics Road, Vancouver, B.C., Canada V6T 1Z2 }
\normalsize

\date{July, 2006}
%\small Notes started July 28, 1998 (old file: prop-adj.tex). 
%Major changes Feb 6, 2001. Last \TeX ed on \today}
\maketitle

\begin{abstract}
\noindent
A class of continuous renormalization group flows with a 
dynamical adjustment of the propagator is introduced and 
studied theoretically for fermionic and bosonic quantum field theories. 
The adjustment allows to include 
self--energy effects nontrivially in the denominator of the propagator
and to adapt  the scale decomposition to a moving singularity,
and hence to define flows of  Fermi surfaces in a natural way. 
These flows require no counterterms, but the counterterms used 
in earlier treatments can be constructed using them. 
The influence of propagator adjustment on the strong--coupling behaviour
of flows is examined for a simple example, and 
some conclusions about the strong coupling behaviour of renormalization
group flows are drawn.
\end{abstract}

\section{Introduction}
The renormalization group (RG) method provides a way to attack,
and in many cases solve, the infrared problem of models in 
field theory, statistical mechanics and many--body theory.
Infrared problems are ubiquitous and physically natural, since they are
inherent to systems with gapless excitations, such as critical systems
of statistical mechanics, massless field theories, and systems with spontaneously
broken continuous symmetries. 

In systems of correlated fermions modelling metals, 
excitations around the Fermi surface are gapless. 
The attempt to treat interaction effects by naive perturbation theory
leads to a severe infrared problem
in the form of divergences due to small denominators. 
The RG has been used for weak, short--range two--body interactions
to cure this problem in perturbation theory and to classify the interaction terms by
their power counting degree \cite{FT,FST}.
The one--body (or self--energy) terms, quadratic in the fermion fields, are the relevant ones, 
followed by the marginally relevant parts of the two--body interaction,
which are quartic in the fermion fields. All higher terms are power counting irrelevant
at weak coupling.
The self--energy terms change the Fermi surface, 
which is the location of the singularity  of the fermion propagator. 
Because the singularity has to be kept track of precisely, 
neglecting these terms leads to problems even in formal perturbation 
theory (provided it is done to higher than second order).
For this reason, the self--energy terms are the most relevant ones. 
The marginally relevant  parts of the two--body terms
determine the quasiparticle interactions and give information 
about ordering tendencies and collective phenomena.
The hierarchy of equations couples the evolution of 
the different terms. 

By the above arguments, a possible way to proceed
is to assume that the Fermi surface given by the free propagator is
already that of the interacting system. 
One can then take the dispersion function of a certain form,  
hence the interacting Fermi surface of a  certain shape, 
assuming that this can be realized in an interacting system,
and treat the remaining selfenergy effects by keeping them 
in the form of two--legged vertices. 
This also involves the assumption that there is no large correction to the 
quasiparticle weight.  Both assumptions can, with considerable technical
effort, be verified for the situation of  smooth, positively curved
Fermi surfaces \cite{FST,PeSa,crg}. 
Technically, this is done in \cite{FST} using counterterms 
to fix the Fermi surface.
The counterterm function determines the shift from the
Fermi surface of the free to that of the interacting system. 
The above realizability question
takes the form of an inversion problem, which is solved under certain 
conditions in \cite{FST}. 

Although the importance of the self--energy terms is  known and 
clearly acknowledged in the literature, a way to proceed in applied 
studies has been to focus on the study of the two--body interactions
while neglecting self--energy terms in the flow. 
The main motivation for this is simplicity. Controlling the 
relevant selfenergy terms carefully is technically complicated 
already in the case of a smooth Fermi surface. 
The counterterm method of \cite{FT,FST} provides a conceptually clear way of 
dealing with the problem, but it has been used only in a few applied 
studies so far.  

In the two--dimensional Hubbard model, 
the case with Van Hove singularities is of interest 
for high-$T_c$ superconductivity \cite{Markiewicz}.
For this case, a renormalized expansion using counterterms
is developed, and some regularity properties of 
the self--energy are shown,  in \cite{SFS}. 
However, the question of  realizability of singular Fermi surface shapes
has not been fully answered yet, although it is very important for 
deciding how generic Van Hove singularities really are in 
two--dimensional interacting systems
\cite{SFS}. 
The effects of Fermi surface dynamics near Van Hove points
can be rather nontrivial, 
since the Fermi surface is most easily moved in the vicinity 
of a zero of the gradient of the dispersion relation. 
This point was first clearly realized in \cite{MetznerFFS},
and it was shown that  quantum fluctuations 
of the Fermi surface can arise and 
lead to non--Fermi liquid behaviour \cite{MetznerdellAnna}.
In this situation, it is useful to have a method in which 
the adjustment of the Fermi surface is done in the flow
itself, because the concept of a non--fluctuating Fermi surface itself
may make sense only down to a certain scale, below which 
the fluctuations emerge and may become dominant. 

The method discussed in this paper 
allows to use dynamically adjusting scale
decompositions and to study Fermi surface flows. 
As a method, it applies more generally and can be used flexibly 
to adapt and adjust propagators in bosonic and fermionic (or mixed) systems,
both for continuous and discrete flows.
Variants of the method are used in mathematical studies 
of Fermi surface flows and Fermi liquid theory \cite{PeSa}. 
Also, the method has already been used to calculate the Fermi surface flow
for the Hubbard model (see the Appendix of \cite{HSFR}).

Various renormalization group schemes have already been used
to study correlated fermion systems: Polchinski's original scheme
\cite{Zanchi}, the Wick ordered scheme \cite{crg,HM},
the 1PI scheme \cite{Wett1PI,SH,HSFR}.
Flow equations for the 2PI functions were introduced \cite{Wett2PI} 
and studied for quasi-1d systems \cite{Dupuis}. 
In this context one may well ask why one would want  to investigate 
another scheme. In the following, I give some further motivations for this. 

As is well--known, all the above--mentioned schemes are equivalent on the level of 
untruncated hierarchies, because all of them contain 
the full information about all Green functions, as long as 
they are kept as infinite hierarchies. 
But they start to differ once the approximations that 
are necessary in nontrivial applications, are made,
most importantly truncations of infinite hierarchies, 
gradient expansions of the generating functionals,
or simplifying assumptions about the dependence of the 
Green functions or vertices on momenta and frequencies. 
It is then useful to understand which features of a given scheme 
make it useful for a particular situation.

All of the above--mentioned schemes include, 
in their untruncated form, all self--energy effects 
in the flow, albeit in a rather different way. In their original form, 
the Polchinski and Wick ordered schemes keep self--energy effects
as vertices, while the 1PI scheme includes them in the  most natural way, 
namely by having full propagators on the lines. 
On the other hand, the Wick ordered scheme can be transformed 
so as to have full propagators on the lines (see Section 4.7.2 of \cite{msbook}),
and a similar transformation also applies to the Polchinski scheme. 
However, the standard momentum space setup of the RG
with a cutoff in momentum space near to the Fermi surface
does not include an adaptive scale decomposition, 
and thus does not allow for the above--mentioned flow of the 
Fermi surface. This is the case in all schemes, so a correct procedure
would require using counterterms  even in the 1PI scheme
(in most applied studies, the Fermi surface shift was neglected,
so that this problem did not appear). 
Alternatively,  one can try not to use a cutoff on the fermion momentum
\cite{Tflow,Kopietz}, but power counting has been 
done rigorously in more than one dimension 
only in the momentum space scheme \cite{FT,FST,crg}.

In two-- or higher--dimensional applications, 
it is a serious practical problem to take the 
momentum and frequency dependence 
of the flowing vertex functions into account. 
The full propagator $G$ of the 1PI scheme has support on all scales 
above the cutoff scale. This makes it necessary to extend the loop integrals
over all of momentum space, which in turn requires having 
an accurate form of $G$, {\em and} of the higher vertex functions,
over a wide range of momenta and frequencies.
This has not been achieved so far. The situation in the other schemes
is better in this respect: 
in the Polchinski equation, all internal lines carry a derivative of the cutoff function.
Therefore, at scale $\Esc{}$, the propagators get evaluated 
only at values of  the momentum $k$ with $|e(k)| \approx \Esc{}$. 
The integration procedure for Polchinski's equation makes the 
equation nonlocal in the scale parameter in the standard way \cite{BK},
but a significant  advantage is that whenever a momentum region $|e(k)| \approx \Esc{}'$ 
further away from the Fermi surface contributes 
to a loop integral in the flow equation, it does so only via a propagator 
on scale near to $\Esc{}'$ , which was already calculated 
in an earlier stage of the flow, and which does not need to be adjusted 
in every further integration step.
In the Wick ordered scheme, all propagators are supported at or below the cutoff, 
so only small neighbourhoods of the Fermi surface occur in the loop integrals
at low scales. Thus the propagators and vertices are needed only close to the 
Fermi surface and at small frequencies.

%The adaptive flow is set up by writing a general form for a scale--dependent 
%generating function and then requiring that the generating function is
%invariant under the RG flow. The standard momentum space flows used in 
%\cite{HSFR,Zanchi,HM} are all of this form, but they include
%restrictions on the covariance in the Gaussian measure that
%are dropped from the setting used here. Instead, the covariance, 
%which corresponds to the propagator for the low--energy particles, 
%is determined by the requirement that certain terms be removed
%from the interaction.  In Section 2, I derive this flow and discuss
%some of its general properties. The resulting equations are such 
%that the existence of a solution is not obvious. A hint at an existence
%proof is given by showing that the continuous flow is the limit
%of a natural discrete flow. The full existence proof follows from 
%\cite{PeSa}. In Section 3, I discuss the example of the Fermi surface
%flow. Finally, Section 4 contains an analysis of a toy model. 

In studies and talks by the G\" ottingen group \cite{Meden}, 
it was pointed out that in simple models, the 1PI scheme 
has a better strong coupling behaviour than the other schemes. 
Namely, the 1PI flow equation is accurate in these toy models 
at much larger coupling values than one might believe at first,
far beyond the weak coupling regime, and it remains accurate
in a region  of intermediate to strong couplings where all other
schemes fail.
The dynamically adjusting scheme is examined for this 
toy model in Section 4. It turns out that propagator 
adjustment, namely putting selfenergy effects into the 
denominator instead of keeping them as vertices, 
is a necessary, but not sufficient, 
ingredient in stabilizing the equations at 
strong coupling. 
The Polchinski scheme still fails at strong
coupling, while the dynamically adjusted Wick ordered scheme 
is accurate up to arbitrarily large couplings. 
The analysis also clarifies the reasons for the success of the 1PI  scheme. 
In Section 4, I also give a more general discussion about strong 
initial couplings, and show that for reasons that are simple,
but specific to the RG method, the ``weak--coupling''  RG flow 
is a well--defined starting point at strong coupling also in ``real'' models
(such as  the two--dimensional Hubbard model),
contrary to what one may believe at first, and to what is stated 
in many places. Whether this can be used efficiently in 
computations depends on certain relative rates of growth; 
this is under investigation and will be discussed at the end.  

\section{Theory} 

\subsection{Setup} \label{secsetup}    
The partition function of a general bosonic or fermionic theory 
with source fields $\Ps$ (the generating function for the Green functions) 
is
\begin{equation}\label{P0def}
P_0(\Ps) = {\cal N}
\int \cD \Th\;
\E^{-{1\over 2} (\Th, A_0 \Th) - \cV_0(\Th) + (\Ps,\Th)}.
\end{equation}
Here the fields $\Th_X$ are indexed by elements $X$ of a finite set $\cX$.
$\cX$ is taken finite to make the generating function mathematically well--defined.
In many applications, $\cX=\Gamma \times I$, 
where $\Gamma$ is a finite space-time lattice and $I$ is a finite set of internal indices,
but this assumption will not be needed for the general theory.
If $\cX$ comes from such a lattice regularization of a continuum theory,
the limit of lattice spacing to zero and volume to infinity can be taken later
in the RG equations (and after a volume factor has been divided out 
in the field--independent term). 
Readers who trust that all this can be done properly (or who don't care)
can also think of a continuous space $\Ga$ right away.

For bosons, the $\Th_X$ are real--valued; for fermions, 
each $\Th_X$ is one of the generators of a Grassmann algebra. Spinors and vector fields are included in this setting
by an appropriate choice of $I$, and complex bosonic fields as well, 
by regarding them as two--component real fields (with the corresponding 
index again part of $I$). 
The commutation relations are
\begin{equation}
\Th_X \Th_Y = \zeta \Th_Y \Th_X 
\end{equation}
with $\zeta=1$ for bosons and $\zeta=-1$ for fermions. 
The integration measure is $\cD \Th = \prod_{X \in \cX} \dd \Th_X$, 
and  $(\Ps,\Th) = \al \sum_{X \in \cX} \Ps_X \Th_X$ with $\al >0$; 
(when considering a lattice approximation, $\al $ is the volume of the unit cell 
of the lattice).
The action of the operator $A_0$ is $(A_0 \Th)_X =  \al \sum_Y (A_0)_{XY} \Th_Y$; 
the matrix elements $(A_0)_{X,Y}$ are in general complex. 
An operator $C$ is called {\em $\ze$--symmetric} if  
\begin{equation}\label{covtyp}
(\Ps, C \Th) =  \zeta (C \Ps, \Th) ,
\end{equation}
%that is, $C^T = \zeta C$, 
(i.e.\ $C_{X,Y} = \zeta C_{Y,X}$). 
In the bosonic case, Re $C$ is also required to have nonnegative eigenvalues, so that
for all $\Th$, 
\begin{equation}
(\Th, (\mbox{ Re } C) \Th ) \ge 0 
\end{equation}
(in the case of complex fields viewed as two--component real fields, 
this corresponds to the usual condition that the hermitian part of the 
covariance is positive definite). 
Assume that the $A_0$ in \Ref{P0def} is $\ze$--symmetric.
If $A_0$ is invertible, both $A_0$ and the {\em propagator} $C=A_0^{-1}$ are $\ze$--symmetric. 
The normalization factor ${\cal N}$ is chosen such that $P_0 (0)=1$ for $\cV=0$. 
For bosons, I also assume that 
$\cV$ is bounded below and that  ${1\over 2} (\Th, A_0 \Th) +\cV(\Th)$
grows faster than linearly in all $|\Th_X|$ for large $|\Th_X|$. 
Under these hypotheses, the integral in \Ref{P0def} is a convergent
finite--dimensional integral in the case of bosons. In the case of fermions, 
it is a linear functional on a  finite--dimensional space. 
The theory can also contain both bosonic and fermionic fields. 
Every formal functional integral can be regularized so that it takes this form.

The logarithm of $P_0$ generates the connected Green functions
of the theory. Assume that $A_0$ is invertible and let $C=A_0^{-1}$.
A shift in the integration variable gives (using the symmetry 
of the quadratic part under $\Theta \to -\Theta$)
\begin{equation}\label{Cgen}
P_0(\Ps) = 
\E^{{1\over 2} (C \Ps, \Ps)}
\int \dd \mu_{C} (\Th)
\E^{ - \cV_0(\Th+C\Ps)}
\end{equation}
with $\dd \mu_{C} (\Th)$ the normalized Gaussian measure with
covariance $C$. In this form, one can take a limit where some of 
the eigenvalues of $C$ go to zero; in the bosonic case this corresponds 
to restricting  the integral to the subspace $\mbox{ker } C$.
Typically, this situation arises by introducing cutoff functions
that vanish strictly in parts of momentum space; see the following examples. 

Although the equations derived in this paper work for the general 
case set up above, it is useful to keep two prototypical examples in mind. 

The first example is  scalar field theory on a finite lattice $\Gamma \subset\bZ^3$
where $|I| = 1$, $(A_0)$ is the discrete Laplacian with (say)
Dirichlet boundary conditions, and 
$\cV_0 (\Th) = \sum_{X \in \Ga} V(\Th_X)$, with $V$ a polynomial
that is bounded below. In that case, $(A_0)_{X,Y} = a_0 (X-Y)$ with 
Fourier transform $\hat a_0 (k) = 2 \sum_{i=1}^3 (1 - \cos k_i)$
for $k \ne 0$. The momentum $k=0$ is removed by the Dirichlet condition;
of course, other boundary conditions can be used, too, provided
the Laplacian is restricted to the complement of its kernel.
For details, see, e.g., Chapter 2 of \cite{msbook}. 

The second example is the regularized partition function 
for spin $1/2$ fermions on 
a finite sublattice $\Lambda$ of $\bZ^d$, $d \ge 1$.
Here $\Gamma = \{0, \frac{\be}{n}, 2\frac{\be}{n}, \ldots , (n-1)\frac{\be}{n}\} \times \Lambda$,
$I = \{1,-1\} \times \{1,-1\}$, with the first index being a charge index and the 
second the spin index. For $X=(c,\xi) \in I \times \Ga$, with $\xi=(\al,\tau,x)$ 
and $X'=(c',\xi')$, with $\xi'=(\al',\tau',x')$, 
\begin{equation}\label{asycov}
(A_0)_{X,X'} = 
\left(
\begin{array}{rr}
0 & \tilde A_0 (\xi,\xi')  \\
- \tilde A_0 (\xi',\xi) & 0
\end{array}
\right)_{c,c'}
\end{equation}
where $\tilde A_0 (\xi,\xi')= \de_{\al,\al'} a_0 (\ta-\ta',x-x')$
%$a_0 (\ta-\ta',x-x') = \del_{\ta,\ta'} +  h(x-x') - \mu \de_{x,x'} $,
%with $\del $ a first order discrete derivative, $h(-x)=\bar h(x)$ and $\mu$ a real parameter.
and the Fourier transform of $a_0$ is of the form 
$\hat a_0 (\omega,k)= \I \omega - e_0(k)$, where the Matsubara frequency 
$\omega$ takes values in odd multiples of  $\pi$ times the temperature $T=\be^{-1}$
and $e_0$ is a real--valued function that 
has a nontrivial zero level set, the Fermi surface $S_0 = \{ k : e_0 (k) =0\}$. 
$\cV_0 (\Ps)$ is an even element of the Grassmann algebra with 
$\cV_0 (0) = 0$. In the simplest case, $\cV_0$ is a quartic polynomial 
in the fields, e.g.\ $\sum_{X_1,\ldots, X_4} v_0(X_1,\ldots, X_4)\; \Ps_{X_1} \ldots \Ps_{X_4}$,
but it can be much more general. 
The number $n$ is an auxiliary quantity coming from an application 
of the Lie--Trotter product formula. To recover a representation of the 
partition function of the many--fermion model, as given by a trace over Fock space, 
one must take $n \to \infty$ first. 
For details, see Chapter 4 of \cite{msbook}. 

\subsection{Renormalization group}\label{motiv}
The renormalization group comes into play when the inverse
of $A_0$ becomes unbounded, or even fails to exist,
in the limit $|\cX| \to \infty$, or when the inverse remains bounded,
but its norm becomes nonuniform in an essential physical parameter. 
In the above examples, the Fourier transform diagonalizes $A_0$
and one can read off that the eigenvalues go to zero in the limit of interest.
In Example 1 above, taking $\Gamma $ to be a cube of sidelength $L$, 
$k_i=n_i 2\pi/L $ with $n_i \in \bN_0$; clearly, $\hat a_0 (k)$ 
goes to zero for infinitely many $n=(n_1,n_2,n_3)$ in that limit,
so that $\hat C = \hat a_0^{-1}$ becomes singular.
Similarly, in Example 2, one gets zero eigenvalues when the temperature 
$T \to 0$, and when $k \in S_0$. At small positive $T$, $\hat C=\hat a_0^{-1}$ 
is of order $1/T$. In both cases, $\hat C$, and thus $C$ itself, 
fails to be square integrable in the limit $T \to 0$, so that integrals 
over a square of a propagator (and higher powers of it) are ill--defined. 
This is not just mathematical pedantry but physically relevant --
if in the case of the many--fermion system, $\hat C$ were square integrable
at $T=0$,  there would be no superconductivity,
because the superconducting gap is driven by the $\log \be$ from 
$\int |\hat C|^2$. Similarly, the singularity of $\hat a_0^{-1}$ in the 
bosonic case leads to anomalous decay exponents of correlation functions.
Hence in the following, 
a guideline for getting a useful RG equation in these systems
is that  {\em squares of unregularized propagators are to be avoided}.

A standard approach to obtain a renormalization group equation
\cite{crg} in the above examples
is as follows. Introduce a scale parameter $s \ge 0$, an $s$--dependent energy,
say $\epsilon_s = \epsilon_0 \E^{-s}$, which goes to zero in the limit
$s \to \infty$. Choose a fixed decreasing $C^\infty$ 
cutoff function $\chi_<$ with $\chi_< (x) =1$ 
for $x \le 1/4$ and $\chi_<(x) = 0$ for $x \ge 1$,
and set $\chi_> = 1 - \chi_<$, so that $\chi_< + \chi_> = 1$ is a smooth
partition of unity.  
Decompose the propagator into an infrared part $D_s$
and a part supported at higher scales, $C_s$, via
\begin{equation}\label{decompo}
\hat C(k) 
=
\hat C_s(k) + \hat D_s(k) =
\chi_> \left(\frac{|\hat a_0 (k)|^2}{\epsilon_s^2}\right) 
\hat C (k) + 
\chi_< \left(\frac{|\hat a_0 (k)|^2}{\epsilon_s^2}\right) 
\hat C (k).
\end{equation}
and define the effective action 
\begin{equation}
\cG_s(\Phi) = - \log \int \dd \mu_{C_s} (\Th) \E^{ - \cV(\Th+\Phi)}.
\end{equation}
The partition function is then expressed in terms of the 
effective action as
\begin{equation}\label{oldflow}
P_0(\Ps) = 
\E^{{1\over 2} (C\Ps, \Ps)}
\int \dd \mu_{D_s} (\Th) \E^{ -\cG_s(\Th+\Phi)}.
\end{equation}
The effective action $\cG_s$ is a function of a fields which,
when expanded in the fields, has a quadratic term.
In infrared problems, this term is relevant in the RG sense. 
That is, the order of the singularity may get changed,
producing the anomalous exponents, in the scalar field theory example, 
or, in the many--fermion system,
the changes to the quadratic term also shift the location of the singularity 
of the propagator, the Fermi surface, in Fourier space. 
The above scale decomposition does not take this
into account, because it zooms in on the fixed Fermi surface $S_0$.
Therefore, with such a fixed choice of scale decomposition, 
counterterms are needed to avoid divergences \cite{FST}. 
It is possible to justify the use of counterterms
by an inversion theorem \cite{FST}, but it seems desirable
to have a simple method for moving the quadratic terms from 
$\cG_s$ into the propagator, in a way that the shift of the
singularity gets included in the flow automatically. 
The flow introduced in this paper allows to put  self--energy terms into the propagator
and hence to define a convenient adaptive scale decomposition both in continous
(and in discrete) renormalization group flows. 
It is such that an analogue of \Ref{oldflow} continues to hold, but with a propagator
that changes dynamically in the flow, and $\cG(s) $ replaced by an 
interaction from which all, or part of, the quadratic part is removed.

Finding a scheme that implements an adjusting scale decomposition 
in a differential equation is not completely straightforward.
On graphical grounds, one expects the full propagator to be 
$G= (A - \Si)^{-1}$, where $\Si$ is the Dyson self--energy. 
Thus the first idea may be to make the ansatz 
\begin{equation}
C(s) = A(s)^{-1} \; 
\ch_> \left(\frac{A(s)^*A(s)}{\epsilon_s^2}\right)
\end{equation} 
with $A(s) = A_0 - \Si(s)$, $\Si(s) $ to be determined.
$C(s)$ is well-defined
because $\ch_>$ is nonzero only on a subspace where $A_0 -\Si(s)$
can be inverted, and the cutoff function now adapts dynamically
because it depends on $\Si(s)$.
The propagator of the unintegrated fields must be defined as
$D(s) =C - C(s)$  to ensure that $P(s,\Ps) = P_0(\Ps)$.
But this has a singularity 
at the noninteracting Fermi surface for any $s$, 
as well as one developing  on the new one,
so (apart from being a very unnatural object) it does not vanish 
for $s \to \infty$, so that not all degrees of freedom get integrated out.
Then this flow would not give useful information about the model, 
even if it converged for $s \to \infty$. 
Trying instead to keep the form of \Ref{oldflow}, but putting
\begin{equation}\label{problema}
D(s) = A(s)^{-1} \; 
\ch_< \left(\frac{A(s)^*A(s)}{\epsilon_s^2}\right)
\end{equation}
leads to disaster: The derivative $\frac{\del D(s)}{\del s}$ contains 
a term $- A(s)^{-1} \frac{\del \Si(s)}{\del s}  A(s)^{-1} \ch_< $, 
which has a square of the inverse, {\em but no infrared cutoff}
because $\chi_< (x) =1$ for $x \le 1/4$. 
In all cases where the self--energy shifts the Fermi surface, 
$\frac{\del \Si(s)}{\del s} \ne 0$ on the singularity set, 
so that effectively, a square of the propagator appears. 
Because this propagator is not square integrable, this scheme leads to divergent 
integrals already in the RG differential equation at fixed $s$.
The method developed in the following sections avoids these problems. 

\subsection{Conditions on the adaptive flow}
I now return to the general field theoretic situation of Section \ref{secsetup}
and consider the generating function in the form \Ref{Cgen}, i.e.\
\begin{equation}\label{D0gen}
P_0(\Ps) = 
\E^{{1\over 2} (D_0 \Ps, \Ps)}
\int \dd \mu_{D_0} (\Th)
\E^{ - \cV_0(\Th+D_0\Ps)}
\end{equation}
Instead of $D_0 = C=A_0^{-1}$ I make the slightly more general  choice
\begin{equation}\label{D0def}
D_0 = A_0^{-1} \chi_0
\end{equation}
where $\chi_0$ can be chosen as an operator that cuts off very large energies,
chosen such that $D_0$ is $\zeta$--symmetric,
and for bosons, nonnegative (but $\chi_0 =1$ is also allowed). 
In applications, such an ultraviolet cutoff allows to start the 
flow at a scale where degrees of freedom with very high energies have already been 
integrated over.
The flow is set up by posing restrictions on the general, $s$--dependent
partition function
\begin{equation}\label{newflow}
P(s,\Ps) = \E^{K(s) + \frac12 (G(s) \Ps,\Ps)} \int d\mu_{D(s)} (\Th) 
\E^{- \cV(s,\Th+ S(s) \Ps)} .
\end{equation}
Here $K$, $G$, $\cV(\cdot,\Ps)$, $S$, {\em and} $D$ to be determined as functions of $s$.
For fixed $s$, $K(s)$ is a complex number, $S$ is an operator, $G$ and $D$ are 
$\ze$--symmetric operators, and $\cV(s,\Ps)$ is a function of the fields $\Ps$, 
which, for bosons, has to be such that the integral exists (see below).

I impose the following general conditions.  

\begin{enumerate}

\item[\Heins]
$K(0)=0$, $G(0)=D(0)=S(0) = D_0$, and $\cV(0,\psi) = V_0 (\psi)$.

\item[\Hzwei]
For all $s\ge 0$ and all $\Ps$, $\; P(s,\Ps)=P(0,\Ps)$. 

\item[\Hdrei]
 $D(s) \to 0 $ as $s\to \infty$.

\end{enumerate}
These conditions have the following simple interpretation. 

\begin{enumerate}

\item
$P(0,\Ps)=P_0(\Ps)$ is the generating function of the model of interest. 

\item
The generating function $P$, and hence all correlation functions, 
are invariant under the RG.

\item
In the end, all degrees of freedom are integrated out.

\end{enumerate}

\noindent
The field--independent term $-K(s)$ is (up to a factor $\be^{-1}$)
equal to the free energy.
The term involving $G (s) $ naturally arises once one starts
adding terms to the denominator of the propagator. 
This is most easily seen in the case when one requires that $\cV(s,\Ps)$ 
should contain no quadratic term in $\Ps$. 
Then, if the limit $s \to \infty$ exists and $D(s) \to 0$ in that limit
(as required by condition 3),
\begin{equation}
P(\infty,\psi) = \E^{K_\infty + \frac12 ( G_\infty \Ps,\Ps)} \; 
\E^{-\cV_\infty (S_\infty \Ps)} .
\end{equation}
Because $\cV(s)$ has no quadratic term for any $s$, the same holds 
for $\cV_\infty = \lim\limits_{s\to\infty} \cV (s)$. Therefore the term involving
$G_\infty$ is the only quadratic term and hence $G_\infty$ is the full propagator. 
The limit exists e.g.\ if the temperature is above all critical temperatures
\cite{crg,DR,PeSa} or if the Fermi surface is asymmetric under reflections
\cite{FKT}.

Thus, the full propagator $G$ is built up gradually in the RG flow.
For finite $s$, \Ref{newflow} has the usual RG interpretation that 
$D(s) $ is the propagator of particles with interaction $\cV(s)$,
but now $\cV(s)$ can be chosen not to contain any self--energy terms.
Thus, even if the limit $s\to \infty$ cannot be taken because the flow
is accurate only up to some $s_0$ (see, e.g.\ \cite{HM,HSFR}), 
the effective propagator and
interaction can still be used as an input for the theory of the 
excitations with energy below $\epsilon_{s_0}$.

The above conditions do not determine $K(s)$ and the functions
$G(s)$, $\cV(s,\Ps)$, $S(s)$, and $D(s)$ uniquely.
They are just the minimal conditions one wants to impose. 
The freedom of choosing the various functions can be used
to impose the constraint that the quadratic part of
$\cV(s)$ should vanish, or that a certain part of it be removed
from $\cV(s)$ and put into $D(s)$ instead. 

The flow \Ref{oldflow} satisfies \Heins--\Hdrei, 
but it restricts to the propagator $D(s) = \chi_s C$ 
with the fixed scale function $\chi_s$ of \Ref{decompo}, 
which does not adjust to a shifting location of the singularity. 
This condition is relaxed when one drops 
\Ref{oldflow} in favour of \Ref{newflow}. 
In particular, one can choose the functions such 
that the propagator $D(s)$ incorporates all, or part of, the 
selfenergy corrections. In applications, this means that one can 
directly study the flow of the Fermi surface, the flow of 
superconducting gaps and the like, without having to use 
counterterms.

\subsection{The RG differential equations}\label{rgdesec}
I seek only  solutions to the conditions \Heins--\Hdrei\
for which $K$, $S,G,D$ and $\cV$ are differentiable in $s$.
Then \Heins\ and \Hzwei\ are equivalent to the initial--value problem
\begin{eqnarray}\label{ivp1}
\frac{\del}{\del s} P (s,\eta) 
&=&
0
\\
K(0)=0, \quad G(0)=D(0)=S(0) 
&=& 
D_0, \quad \cV(0,\psi) = \cV_0 (\psi),
\nonumber
\end{eqnarray}
from which the RG differential equation will be obtained.

For a  $\ze$--symmetric operator $A$, the Laplacian 
in field space is defined as
\begin{equation}
\Delta_A = ( \de_\Ps \; , \; A  \;\de_\Ps ) 
\end{equation}
where $\de_\Ps = \al^{-1} \frac{\del}{\del \Ps}$.

\begin{satz}\label{th1}
Let $s \mapsto Q(s)$ be given, with $Q(s)$ $\zeta$--symmetric 
for all $s$ and bounded for all $s$. 
Assume that there is a solution $(A,\chi)$ to the initial--value problem
\begin{equation}\label{ivp2}
\dot A (s) 
=
- \chi(s) Q(s) , \quad
A(0) = A_0, \quad
\chi(0) = \chi_0,
\end{equation}
such that $s \mapsto \chi(s)$ is differentiable and
the low--energy propagator
\begin{equation}\label{myway}
D(s)
=
A(s)^{-1} \chi(s) 
\end{equation}
and the fluctuation propagator
\begin{equation}\label{Fsdef}
F(s)
=
A(s)^{-1} \frac{\del}{\del s}\chi(s) 
\end{equation}
exist and are $\ze$--symmetric, and that $D(s)$ is nonnegative for bosons.
Assume that $K$ and $\cV$ satisfy the initial--value problem 
\begin{eqnarray}\label{mnseq}
\dot \cV(s,\Ps)
&=&
\E^{\cV(s,\Ps)} \frac12 \Delta_{F(s)} \E^{-\cV(s,\Ps)}
+
\frac12 (Q(s) \Ps,\Ps) 
+ \dot K - \frac{\zeta}{2} \Tr (DQ) ,
\nonumber\\
\cV(0,\Psi) 
&=& 
\cV_0(\Psi), \qquad
K(0)=0
\end{eqnarray}
(thus in particular $\cV(s,\Psi)$ is differentiable in $s$ and 
at least twice differentiable in $\Psi$)
and the condition that $\cV(s,0) = 0$ for all $s$.
Set 
\begin{equation}\label{Ssdef}
S(s) = A(s)^{-1} \chi_0
\end{equation}
and let $G$ be the solution to 
\begin{equation}
\dot G (s) 
=
S(s)^T Q(s) S(s), \quad
G(0) = D_0 .
\end{equation}
Then the functions $K,G,D,S$ and $\cV$ solve the initial value problem \Ref{ivp1},
hence they satisfy \Heins\ and \Hzwei.
\end{satz}

\noindent
This theorem is proven in the next subsection.

\bigskip\noindent
{\bf Remarks. }

\begin{enumerate}

\item
$D$ must be $\ze$--symmetric, and for bosons it must be nonnegative. 
A  way to a\-chieve this is to restrict to  $\ze$--symmetric $A(s)$
that commutes with its adjoint $A^* (s)$, 
and to take $\chi(s)$ as a function of $A^*A(s)$, so that the two commute,
e.g.\ $\chi(s) = \chi_< \left(\frac{A(s)^* A(s)}{\epsilon_s^2}\right)$. 
The conditions on $A$ are satisfied in the above examples 
and in most other interesting applications. 

\item
The dependence of the cutoff function $\chi(s)$ on $A(s)$ allows for an 
$s$--depen\-dent, adaptive scale decomposition. This makes 
showing the existence of a solution to \Ref{ivp2} nontrivial. 
This existence problem will be discussed in more detail 
in Section \ref{sec:2.9}.
 
\item
The fluctuation propagator $F(s)$ given in \Ref{Fsdef}
is now indeed proportional to the derivative of the scale function,
thus has an infrared cutoff. Therefore the problem mentioned 
after \Ref{problema} is absent from this flow. 

\item
The equation for the full propagator $G$ can be rewritten as
\begin{equation}\label{Geek}
G(s) = A(s)^{-1} 
-
\int_0^s \dd t \; 
A(t)^{-1} (1-\chi(t)) Q(t) A(t)^{-1} .
\end{equation}
Thus in general, $G(s) \ne A(s)^{-1}$, also in the limit as $s \to \infty$. 
This must be so for general reasons, as will be explained 
in Section \ref{diagrammar}.

\item
The strategy to satisfy \Hdrei\ as well is to choose
a solution for which the support of 
$\chi(s)$ becomes empty as $s \to \infty$,
and for which the singularity of $A(s)^{-1}$ as $s\to \infty$ 
is such that $D(s) = A(s)^{-1} \chi(s)$ still vanishes in the limit. 

\item
Theorem \ref{th1} gives a family of  solutions to the conditions \Heins\ 
and \Hzwei\ para\-me\-tri\-zed by $Q$. 
Moreover, it will become clear in the proof that even given a fixed $Q$, 
there are many other possibilities to choose the equation for $\cV$, 
and as explained, there is a further freedom of choosing $\chi$ 
as a function of $A$ when seeking a solution $(\chi,A)$ to \Ref{ivp2}.
Thus, there seems to be an enormous freedom in solving \Heins--\Hzwei.
However, not every choice has good analytical properties, 
and most solutions diverge already at a small value of $s$. 
To have a bounded solution on a maximal interval for $s$
requires a careful choice of $Q(s)$. For instance, 
in the Fermi surface problem in $d\ge 2$, 
$Q$ must take the Fermi surface
shift into account fully; whether it also needs to contain
corrections to the Fermi velocity and the coefficient of $\omega$
depends on finer details, such as the shape of the Fermi surface. 
If this is done properly, the limit $s \to \infty$ of the flow exists
at temperatures above the mean--field transition temperature
\cite{PeSa}, and \Hdrei\ is satisfied. 

\item
Compared to the Polchinski equation obtained 
from \Ref{oldflow} by differentiating with respect to $s$, which reads 
\begin{equation}\label{boltschi}
\dot \cG_s = \E^{\cG_s} \frac12 \Delta_{\dot D_s} \E^{-\cG_s} ,
\end{equation}
(with $D_s = C \chi_< $ as in \Ref{decompo}), there are two changes. 
First, there are terms of order 0 and 2 in the fields $\Ps$ on the right hand side of 
\Ref{mnseq}, which can be chosen 
to cancel the field--independent and quadratic part arising from 
the first term on the right hand side, so that $\dot \cV(s)$ 
starts with terms of order $4$ in $\Psi$.
Second, the fluctuation propagator $F(s)$ now
includes selfenergy corrections from the integration
up to scale $s$ (if $Q(s)$ is chosen to absorb only
selected parts of the quadratic part generated by the integration,
e.g.\ only cancelling the Fermi surface shift, then 
the propagator $A(s)$, and hence $F(s)$, 
contain only those).

Thus \Ref{mnseq} has a simple and natural interpretation:
the propagator is allowed to change dynamically, and the selfenergy
(or part of it) can be put into the denominator.
Although the RHS of \Ref{boltschi} has the same structure
as the first term on the RHS of \Ref{mnseq}, the change
of the quadratic part, and the ensuing 
replacement of $\dot C_s$ by $F(s)$ leads to a major change in the 
behaviour of the RG equations. Some aspects will be discussed 
in Section \ref{toysec}.

\end{enumerate}

\subsection{Proof of Theorem 2.1} 

To make  the proof more readable, 
I first give an algebraic derivation which applies both to 
bosons and fermions, and in a subsequent section provide the 
mathematical justifications of all steps. 

\subsubsection{Derivation of the RG equations}
Define the convolution with the Gaussian measure as
\begin{equation}
(\mu_D * F) (\Phi)
=
\int \dd \mu_D (\Theta ) F(\Theta + \Phi)
\end{equation}
%If $F$ is a polynomial in the fields, then \cite{msbook}
%\begin{equation}
%\left( \mu_{D(s)} * F \right) (\Ps)
%=
%\E^{\frac12 \Lap_{D(s)} } F (\Ps).
%\end{equation}
The partition function is 
\begin{equation}
P(s,\eta) 
=
\E^{K(s) + \frac12 (G(s) \eta , \eta) }
\; 
\left(
\mu_{D(s)}
* \E^{-\cV(s) }
\right)
(S(s) \eta)
\end{equation}
If $W$ is differentiable in $s$ and $\Psi$,
\begin{equation}
\frac{\dd}{\dd s} W(s, S(s) \eta)
=
\left\lbrack
\frac{\del W}{\del s} (s,\Psi)
+
\cD_{H(s)} W(s,\Psi) 
\right\rbrack_{\Psi=S(s) \eta}
\end{equation}
with 
\begin{equation}
H(s) = \dot S(s) \, S(s)^{-1}
\mbox{ and }
\cD_H 
=
\left(
H\, \Psi \, , \, \de_\Ps         %\frac{\de}{\de \Psi}
\right)  .
\end{equation}
The equation $\frac{\dd P}{\dd s} = 0 $ implies
\begin{equation}\label{goobo}
\left(
\dot K(s) + \frac12 (Q(s) \Psi, \Psi) + \cD_{H(s)} + \frac{\del}{\del s}
\right)
\left( \mu_{D(s)} * \E^{-\cV(s)} \right) (\Ps)
%\E^{\frac12 \Lap_{D(s)} } \E^{-\cV(s,\Psi)} 
= 0
\end{equation}
with 
\begin{equation}
Q (s) 
=
\left( S(s)^{-1}\right)^T\; \dot G (s) S(s)^{-1} . 
\end{equation}
Using  %(see, e.g.\ \cite{msbook})
\begin{equation}\label{Lappi}
\left( \mu_{D(s)} *  \E^{-\cV(s,\Psi)}\right) (\Ps)
=
\E^{\frac12 \Lap_{D(s)} }  \E^{-\cV(s,\Psi)} (\Ps)
\end{equation}
gives
\begin{equation}\label{fastsoweit}
\left(
\dot K(s) + \frac12 (Q(s) \Psi, \Psi) + \cD_{H(s)} + \frac{\del}{\del s}
\right)
%\left( \mu_{D(s)} * \E^{-\cV(s)} \right) (\Ps)
\E^{\frac12 \Lap_{D(s)} } \E^{-\cV(s,\Psi)} 
= 0
\end{equation}
Acting with $\E^{\cV(s,\Psi)} \E^{-\frac12 \Lap_{D(s)}}$ on \Ref{fastsoweit}
and using 
\begin{eqnarray}\label{commeq1}
\E^{-\frac12\Lap_D} \cD_H \E^{\frac12\Lap_D}
&=&
\cD_H 
- 
\frac12 \Lap_{HD+DH^T}
\\
\E^{-\frac12\Lap_D} \frac12 (Q\Psi,\Psi) \E^{\frac12\Lap_D}
&=&
\frac12 (Q\Psi,\Psi)
- \frac{\ze}{2} \Tr (DQ)
+ \frac12 \Lap_{DQD}
- \cD_{DQ}
\label{commeq2}
\end{eqnarray}
(recall $\ze=1$ for bosons and $\ze=-1$ for fermions)
I get
\begin{equation}
\dot V 
=
\E^{V}
\left(
\dot K -  \frac{\ze}{2} \Tr  (DQ) 
+
\frac12 (Q\Psi,\Psi)
+
\frac12 \Lap_{F} + \cD_B 
\right)
\E^{-V}
\end{equation}
with
\begin{equation}
F=\dot D - HD -DH^T + DQD
, \quad
B=H - DQ .
\end{equation}
To simplify the equations, I remove the dilation term $\cD_B$, i.e.\
choose $S$ so that
\begin{equation}\label{eq:dodo}
H = S^{-1} \dot S 
=
DQ
\end{equation}
This implies $F=\dot D - DQD$, so that
\begin{eqnarray}\label{eq:40}
\dot V (\Ps)
&=&
\dot K - \frac{\ze}{2}  \Tr  (DQ)
+
\frac12 (Q\Psi,\Psi)
+
\E^{\cV(\Ps)} \frac12 \Lap_{\dot D - DQD} \E^{-\cV(\Ps)}
\end{eqnarray}
This equation makes it natural to regard $Q(s)$ as an input, 
and to seek a solution for $D$,$\cV$, and $K$. Once this is done, 
the remaining functions are obtained by integrating the equations
$\dot S=DQ\; S$ and $\dot G = S^T Q S$.

Let $D$ be of the form \Ref{myway}
with $A(s)$ and $\chi(s)$ the solution to \Ref{ivp2}.
Differentiation with respect to $s$ implies that $F(s)$ %in \Ref{mnseq}
is indeed given by \Ref{Fsdef}. 
With the choice \Ref{myway}, the amputation operator $S$ 
can now also be determined: $\dot S = DQ S = -  \dot A \, A^{-1} \, S$, 
with initial condition $S(0) = A(0)^{-1} \chi_0$ 
has the unique solution $S(s)=A(s)^{-1}  \chi_0$ stated in \Ref{Ssdef}.

Thus Theorem \ref{th1} is proven algebraically for bosons and fermions.
Adapting the derivation to the case of a theory involving bosons and fermions
is straightforward,  
provided that the interaction $V_0$ is an even element of 
the Grassmann algebra. 

\subsubsection{Mathematical remarks}
In the above derivation, it was assumed 

\begin{enumerate}

\item 
that $\cV$ and $P$ are indeed differentiable in $s$ and in $\Ps$

\item
that \Ref{Lappi} holds, and

\item 
that the commutator equations hold, when applied to $\E^{-\cV}$. 

\end{enumerate}
Item 1 follows from the existence of the solution to the initial value 
problem for $K$ and $\cV$, as stated in the hypotheses of  Theorem \ref{th1}
(and already explained there).

Concerning item 2, note that for any polynomial $\Pi$ in the fields, 
\begin{equation}
\left( \mu_{D(s)} * \Pi \right) (\Ps)
=
\E^{\frac12 \Lap_{D(s)} } \Pi(\Ps)
\end{equation}
was proven in \cite{msbook} both for bosons and fermions, 
but $\E^{-\cV}$ is in general not a polynomial. 
Similarly, the commutator equations \Ref{commeq1} and \Ref{commeq2}
hold when applied to polynomials in $\Ps$, but again, the above 
proof involves an application to  $\E^{-\cV}$.

Consider first the case of fermions. In the present setup, 
the Grassmann algebra is finite--dimen\-sio\-nal, 
so every function of the Grassmann variables is a polynomial. 
Thus $\cV (s,\Th)$ and $\E^{\pm \cV(s,\Th)}$ are polynomials in $\Th$
and $P(s,\Ps)$ is a polynomial,  hence infinitely differentiable in $\Ps$. 
Thus \Ref{Lappi} holds, and \Ref{commeq1} and \Ref{commeq2} 
hold as well. 
Moreover $\chi $ is differentiable in $s$
by hypothesis of Theorem \ref{th1}, and $A$ is differentiable in $s$
because it solves the differential equation \Ref{ivp2}, so 
$D$ is differentiable in $s$ as well. By \Ref{Lappi},
the convolution in \Ref{newflow} is a polynomial in $D(s)$
and in $\cV(s,\cdot )$. Thus by expanding $\cV$ in homogeneous parts
of degree $2m$ in $\Psi$ (see also Section \ref{diagrammar}), one easily proves 
by induction on $m$ that $\cV$, and hence also $P$, is differentiable in $s$. 
 
For bosons, $\E^{-\cV(s,\Ps)}$ is not a polynomial, but, under our 
assumptions, it has a fast decay for $|\Psi| \to \infty$. 
This decay and some analyticity
could be used to justify first truncating 
$\E^{-\cV}$  to a polynomial and then taking a limit. 
A different, easier way to adapt the argument is to avoid \Ref{Lappi},
i.e.\ go back to \Ref{goobo} and use 
integration by parts--arguments of the type 
\begin{equation}
\Ps \int \dd \mu_D(\Ps-\Th) \;   F(\Th)
=
\int \dd \mu_d(\Ps-\Th) \; 
\left(
\Th - D \frac{\del}{\del \Th}
\right) \;
F(\Th)
\end{equation}
to bring the equation into the form
\begin{equation}
\int \dd \mu_{D(s)} (\Ps - \Th) \; \cF (s,\Th) = 0
\end{equation}
where $\cF(s,\Psi) $ is the difference between the left and right hand sides
of \Ref{eq:40}.
Keeping $D(s) > 0$ in the derivation and obtaining the general case
of nonnegative $D$ by a limit, an equation of this form implies that 
$\cF (s,\Th) =0$. This yields the same differential equation as the 
algebraic method of the previous section and thus completes the 
derivation for bosons. 

In conclusion, all steps in the derivation of the RG equations in the previous
subsection are justified mathematically. In the case of fermions, 
the smoothness of $\cV (s,\Psi)$ in $s$ and $\Psi$ follows from 
easy general arguments. For bosons, it is part of showing that the 
solution to the equations for $\cV$ and $K$ exists. 

\subsection{Structure of the dynRG hierarchy}\label{diagrammar}

\begin{figure}
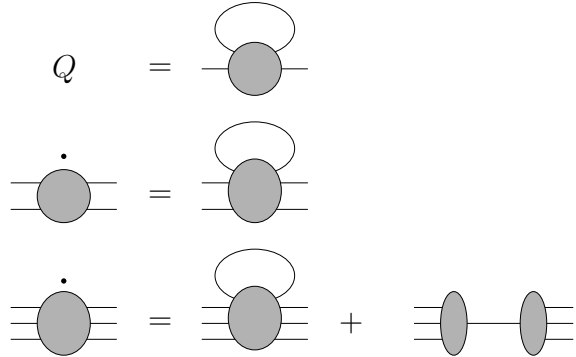

\begin{center}
\figplace{fig1}{0.0 in}{0.0 in}
\end{center}
\caption{The first three equations of the dynRG hierarchy. 
The first equation determines $Q$, which enters the right hand side of 
the differential equation for the inverse propagator. 
The dots denote derivatives with respect 
to the scale parameter $s$.
All internal lines carry a fluctuation propagator $F$.}
\label{fig1}
\end{figure}

For brevity I call the RG differential equation for $\cV, K$ and $A,\chi$
derived above the dynamical RG or dynRG equation. Expansion 
of $\cV$ in powers of the fields gives a hierarchy of equations
for the $m$--point functions. 
In this section, I assume that $\cV$ is even and choose $Q$ and $K$
such that $\cV$ does not contain any quadratic and constant part, 
which gives the hierarchy with the simplest graphical structure. 
The expansion in the fields
\begin{equation}\label{homexp}
\cV(\Psi ) = \sum_{m \ge 2} \cV^{(m)} (\Psi)
\end{equation}
with $\cV^{(m)}$ homogeneous of degree $2m$ in $\Psi$,
leads to the hierarchy 
(shown graphically  for $Q$, $\cV^{(2)}$ and $\cV^{(3)}$ in Figure \ref{fig1})
\begin{eqnarray}
&&
{\dot \cV}^{(m)} 
=
\sum_{n=2}^{m-1} 
\frac12 
(\de_\Ps \cV^{(n)}, \; F \;\de_\Ps \cV^{(m+1-n)})
%\left(
%\frac{\delta \cV^{(n)}}{\delta \Psi}, \; F \, 
%\frac{\delta \cV^{(m+1-n)}}{\delta \Psi}
%\right)
-
\frac12
\Lap_F \, \cV^{(m+1)},  
\quad m\ge 2
\label{Vhi}
\\
&&
(Q(s) \Ps,\Ps)
=
\Delta_{F(s)} \; \cV^{(2)} (s,\Ps)
\label{VhiQ}
\\
&&
\dot K 
=
\frac{\ze}{2} \Tr (DQ)
\label{VhiK}
\end{eqnarray}
By \Ref{ivp2}, \Ref{VhiK} reads
$\dot K 
=
- \frac{\ze}{2}  \Tr  (A^{-1} \dot A)
=
- \sfrac{\del}{\del s}\,
\frac{\ze}{2}
\Tr 
\log A ,
$
so it integrates to 
\begin{equation}
K(s) 
=
- \frac{\ze}{2} 
\left(
\Tr \log A(s)
-
\Tr \log A(0)
\right)
\end{equation}
For fermions, $\ze = -1$ and $\Tr \log A = \mbox{ Pf } A$, 
where Pf $(A)$ denotes the Pfaffian of the antisymmetric matrix $A$. 
For covariances of the form \Ref{asycov}, Pf $(A) = (\det \tilde A)^2$, 
so that 
\begin{equation}
\E^{K(s)} = \frac{\det \tilde A(s)}{\det \tilde A(0)}
\end{equation}
For bosons, $\ze =1$ and $\Tr \log A = \det A$. Thus for real bosons,
\begin{equation}
\E^{K(s)} = \left(\frac{\det A(s)}{\det A(0)}\right)^{-1/2}
\end{equation}
For complex bosons with a hermitian covariance there is a factor $2$ 
which squares this to $\det A(0)/\det A(s)$. 

Thus the relation of the operator $A$ to the free energy density is simple: 
if $A(s) \to A$ for $s \to \infty$, the free energy is (modulo a conventional
factor of the inverse temperature)  simply given by 
\begin{equation}
K =  -\frac{\ze}{2}  \left( \Tr \log  A  - \Tr \log A_0 \right)
\end{equation}
This also clarifies why the equation \Ref{Geek} for $G$ cannot 
simply be $G=A^{-1}$: the exact relation between the full propagator $G$
and the free energy involves the Luttinger--Ward functional. 
It is a nice feature of the dynRG flow that the relation between 
$A$ and the free energy $K$ is so simple.

Clearly, the diagrammatic structure of the hierarchy (see Figure \ref{fig1})
differs from the Polchinski hierarchy obtained from \Ref{boltschi} by the 
absence of  one--particle functions (or two--point functions)
which get absorbed into $A$  via the equation \Ref{VhiQ} for $Q$ and \Ref{ivp2}. 

The standard iterative solution of this hierarchy produces a graphical  expansion.
It should be noted that from a purely graphical perspective, 
the dynamical propagator adjustment does not make any of the functions 
one--irreducible. This is obvious for the six--point and higher functions, 
but also holds for the graphs contributing to the four--point function $\cV^{(2)}$
and to $Q$: taking the tree diagram of the 
six--point function and joining two legs of one of the vertices to a tadpole
produces a one--reducible contribution to $\cV^{(2)}$, 
taking two legs of the other vertex and joining them to a tadpole
gives a one--reducible contribution to $Q$.
The point here is that these two--point insertions 
(the {\em single-scale insertions} defined in \cite{FST}) 
are harmless because they do not generate singularities. 
In the Fermi surface example, the size of such an insertion at scale $s$ is
by power counting (see \cite{crg}) of order $\epsilon_s \sim \E^{-s}$, 
while the size of the propagators on the adjoining lines is of order 
$\epsilon_{s}^{-1} \sim \E^{s}$.
Thus the combination of two propagators
and such a two--point insertion is of order $\E^{-s} \E^{2s} = \E^s$, 
which is the same size as a single propagator. 
However, if one inserted a two--point graph that is integrated from $0$ to $s$, 
its value would be $\sim \int_0^s \dd s' \E^{-s'} = O(1)$. 
Then the combination of this insertion with two propagators is of 
size $\E^{2s}$, hence $\E^s$ larger than it should be. It is contributions
of this type that lead to a problem with unrenormalized expansions. 
Here, this problem does not arise because renormalization is implemented
by a changing propagator. 
In summary, for the analytic questions, namely which terms are large and which are
small, the reducible graphs contributing to $\cV^{(2)}$ and $Q$ are 
inessential.  

\subsection{Wick ordering and self--consistency equations}
Expanding in Wick ordered polynomials provides essential 
simplifications in the mathematical analysis of the RG equations \cite{crg,msbook};
moreover the Wick ordered scheme has been useful for calculations
in two--dimensional systems \cite{HM,RoheMetzner}.
A distinctive feature of the Wick ordered equation is that all propagators
are supported on scales below $\Esc{s}$, so that only low--energy degrees
of freedom enter the integration. 
In the many--fermion example, this means that all momentum integrals 
are restricted to a neighbourhood of the Fermi surface. 
Therefore projections
to the Fermi surface, which are needed to solve the equations numerically, 
are better under control than in other schemes.
An further advantage is
that structures of overlapping loops \cite{FST,crg,S3} become explicit in the 
equations and make the study of improvements to power counting much easier.
These improvements are needed for showing Fermi liquid behaviour in higher 
dimensions and controlling two--particle interactions. 

A natural (but not the only) way to fix the Wick ordering covariance
is to require that tadpoles be removed. In the case of the fixed scale decomposition
\Ref{decompo}, this is easy to do : the condition to cancel tadpoles is to have
$\frac{\del}{\del s} R(s) = -\dot C_s$, because then the linear term in the 
RGDE drops out \cite{crg}. In this case, the solution to that equation is simply
$R(s) = D_s$, with the $D_s$ of \Ref{decompo}. 
In the adaptive scheme \Ref{mnseq}, the corresponding equation is 
\begin{equation}
\frac{\del}{\del s} R(s) 
=
- F(s)
\end{equation}
which is {\em not} solved by $R(s) = D(s)$. The reason for this is
simply that  a Wick ordering that removes all tadpoles requires
information about the propagator at later scales $s' > s$, 
which requires solving  self--consistency equations. 
In theoretical studies, this self--consistency
can be solved by fixed--point arguments. Alternatively,
one can choose the Wick ordering covariance such that it leads to the above
improvements without cancelling tadpoles exactly.

The Wick ordered equation is obtained from \Ref{mnseq}
\begin{eqnarray}
\dot \cV(s,\Ps)
=
&-& \frac12 \Delta_{F(s)} \cV(s,\Ps)
+
\frac12
\left(
\frac{\del \cV(s,\Ps)}{\del \Ps}, 
\;
F(s)
\;
\frac{\del \cV(s,\Ps)}{\del \Ps}
\right)
\nonumber\\
&+&
\frac12 (Q(s) \Ps,\Ps) 
+ \dot K - \frac{\zeta}{2} \Tr (DQ) ,
\end{eqnarray}
by setting
\begin{equation}
\cV(s,\Ps)
=
\E^{-\frac12 \Delta_{R(s)}}
\;
\cW(s,\Ps) .
\end{equation}
Then
\begin{equation}
\dot \cV(s,\Ps)
=
- \frac12 \Delta_{\dot R(s)} \cV(s,\Ps)
+
\E^{-\frac12 \Delta_{R(s)}}
\;
\dot
\cW(s,\Ps)
\end{equation}
and  the resulting equation is
\begin{eqnarray}
\dot
\cW(s,\Ps)
&=&
-\frac12 \Delta_{F(s)-\dot R(s)} \cW (s,\Ps)
\nonumber\\
&+&
\Leval
\E^{\frac12 \Delta^{(1,2)}_{R(s)}}
\frac12  \Delta^{(1,2)}_{F(s)}
 \cW (s,\Ps_1)
 \cW (s,\Ps_2)
\Reval
\\
&+&
\frac12 (Q(s) \Ps,\Ps) 
+ \dot K(s) - \frac{\zeta}{2} \Tr [(D(s)-R(s))Q(s)] .
\nonumber
\end{eqnarray}
Here  the usual combinatorial device \cite{hessel}
of introducing two copies
$\Psi_1$ and $\Psi_2$ of the field has  been used to rewrite 
the quadratic term, and $\Leval \ldots \Reval$ denotes 
evaluation at $\Ps_1=\Ps_2=\Ps$.

The choice of $R$ as the solution to 
\begin{equation}\label{Req}
\dot R(s) 
=
F(s)
\end{equation}
makes the linear term in the equation drop out, as in \cite{crg}.
There are some further conditions on $R$ that were automatically satisfied in the 
case without propagator adjustment in \cite{crg}. The support of $R$ needs to 
go to zero as $s \to \infty$, and $R$ needs to satisfy certain power 
counting bounds to guarantee that Wick ordering does not introduce
singularities. For this reason, rewriting the solution in terms of an 
initial condition as 
\begin{equation}
R(s) = R(0) 
+
\int_0^s \dd t\; 
A(t)^{-1} \dot \chi(t) 
\end{equation}
is not useful. Instead, requiring $R(s) \to 0$  for $s \to \infty$, 
the solution is conveniently written as
\begin{equation}\label{Rinteq}
R(s) 
=
- \int_s^\infty \dd t\; 
A(t)^{-1} \dot \chi(t) 
\end{equation}
However, in contrast to the situation without adjustment of propagators
(where $R(s) = D_s$), this is really a self--consistency equation
because it requires knowledge about the propagator at later RG times. 
For general purposes, such as order--by--order arguments and
power counting, it suffices to know general properties of $R$, 
and to solve for $R$ inductively while solving the RG equations. 
In practical applications, one needs to use approximate solutions 
$\tilde R$ instead of $R$, for which  
$R - \tilde R$ is less singular than $R$ itself.

\subsubsection{The hierarchy of Wick ordered functions}
The expansion 
\begin{equation}
\cW (s,\Ps) 
=
\sum_{m \ge 2} \cW^{(m)} (s, \Ps )
\end{equation}
with $\cW^{(m)}$ homogeneous of order $2m$ in $\Ps$ 
gives the following hierarchy. For $m \ge 2$, 
\begin{eqnarray}
&&
\dot \cW^{(m)} (s,\Ps)
=
- \frac12 \Delta_{F - \dot R} \cW^{(m+1)} (s, \Ps)
\nonumber\\
&+&
\sum_{\ell \ge 0}
\sum_{\mu=2}^{m+\ell -1}
\frac{2^{-\ell -1}}{\ell !}\;
\Leval
(\Delta_R^{(1,2)})^\ell \;
\Delta_F^{(1,2)} \; 
\cW^{(\mu)} (s,\Ps_1)\; 
\cW^{(m+1+\ell-\mu)} (s,\Ps_2)
\Reval
\end{eqnarray}
for $m=1$
\begin{eqnarray}
&&
\frac12 (Q\Ps,\Ps)
=
- \frac12 \Delta_{F - \dot R} \cW^{(2)} (s, \Ps)
\nonumber\\
&+&
\sum_{\ell \ge 2}
\sum_{\mu=2}^{\ell }
\frac{2^{-\ell }}{\ell !}\;
\Leval
(\Delta_R^{(1,2)})^\ell \;
\Delta_F^{(1,2)} \; 
\cW^{(\mu)} (s,\Ps_1)\; 
\cW^{(2+\ell-\mu)} (s,\Ps_2)
\Reval
\end{eqnarray}
and for $m=0$
\begin{eqnarray}
&&
\dot K 
=
\frac{\ze}{2} \Tr [(D-R)Q]
\nonumber\\
&+&
\sum_{\ell \ge 3}
\sum_{\mu=2}^{\ell -1}
\frac{2^{-\ell -1}}{\ell !}\;
\Leval
(\Delta_R^{(1,2)})^\ell \;
\Delta_F^{(1,2)} \; 
\cW^{(\mu)} (s,\Ps_1)\; 
\cW^{(1+\ell-\mu)} (s,\Ps_2)
\Reval
\end{eqnarray}
This hierarchy is shown graphically in Figure \ref{fig2}.
The number of loops of the corresponding graphs is $\ell$. 
As is explicit in the above equation, the contributions quadratic in $\cW$ 
start at two--loop for $\dot Q$ and at three--loop for $\dot K$. 
The choice $F=\dot R$ removes the term that is linear in $\cW$. 
A graphical analysis of the iteration implies that with this choice, 
there are no one--reducible contributions to $\cW^{(2)}$ and to 
$Q$.
Moreover,  if the one--loop term is dropped from 
the eq. for the four-point function, then all graphs contributing to
$\cW^{(2)}$ and $Q$ are 2PI.
%The choice $R=D$ removes the first term in the equation for $\dot  K$
%but leaves the tadpole terms. 

\begin{figure}
\begin{center}
\figplace{fig2}{0.0 in}{0.0 in}
\end{center}
\caption{The first three equations of the Wick ordered dynRG hierarchy, 
with the choice $\dot R = F$ and the truncation that $W^{(m)} = 0 $ for $m \ge 4$. 
In the truncation where also $W^{(3)} = 0$, only the first of the three terms
remains on the right hand side in each equation.}
\label{fig2}
\end{figure}

\subsection{Ladder summations and symmetry breaking}
If the Wick ordering covariance $R$ is chosen as the solution to 
$\dot R = F$ that vanishes for $s \to \infty$, as above, 
and if the hierarchy is truncated by setting $\cW^{(m)} = 0$ for all 
$m \ge 3$, the resulting equation for $\cW^{(2)}$ becomes
\begin{eqnarray}\label{67}
\dot \cW^{(2)} (s,\Ps)
&=&
\frac14\;
\Leval
\Delta_{R(s)}^{(1,2)} \;
\Delta_{F(s)}^{(1,2)} \; 
\cW^{(2)} (s,\Ps_1)\; 
\cW^{(2)} (s,\Ps_2)
\Reval
\nonumber\\
&=&
\frac12\;
\Leval
\left(
\sfrac{\del}{\del s} 
[\Delta_{R(s)}^{(1,2)}]^2
\right)
\cW^{(2)} (s,\Ps_1)\; 
\cW^{(2)} (s,\Ps_2)
\Reval
\end{eqnarray}
The structure of this equation is analogous to that 
obtained by a similar truncation from the 1PI hierarchy
{\em after the 1PI hierarchy has been rearranged in 
Katanin's \cite{Katanin} way}:   Eq.\ (25))
in \cite{SHML} is to quadratic order in the interaction
of the same form as \Ref{67}.  It is also analogous to equations
(4.176) -- (4.178) in \cite{msbook} because there a product
of propagators $D_s \dot D_s$ appears in the flow equation
as well; however, there, $D_s$ does not contain self--energy 
corrections). Note that in contrast to the 1PI case, 
the structure of  the scale derivative of  a square of the Laplacian
(which is essential for the solution of the selfconsistency)
appears in the Wick ordered dynRG in a natural way;
no rearrangement of the hierarchy is necessary.

It is thus clear that, when restricted to a particular channel, 
the equation is easily solved by a geometric series 
in analogy to the procedure in Section 4.5.4. of \cite{msbook}.
In the present framework, restricting to a channel means that 
(possibly after a change of variables), the equation for the 
coefficient function $W_{X,Y,X',Y'}$ of $\cW^{(2)}$ has the 
structure 
\begin{equation}\label{abschan}
\dot 
W_{X,Y,X',Y'} (s)
=
\sum_{U,V,U',V'}
W_{X,Y,U,V} (s)
\dot \Pi_{U,V,U',V'} (s)
W_{U',V',X',Y'} (s)
\end{equation}
which has the structure $\dot W = W  \dot \Pi W$ with the 
product defined in the way evident from \Ref{abschan},
i.e.\ regarding pairs $(U,V)$ as matrix indices. 
Because \Ref{67} contains a derivative of the square of the Laplacian, 
$\dot \Pi$ is indeed explicitly given as the derivative of a matrix $\Pi$.
The solution of \Ref{abschan} is
\begin{equation}\label{wiso}
W(s)
=
W(0) \;
\Big(1- [\Pi (s)-\Pi (0)]\;  W(0)\Big)^{-1} .
\end{equation}
That integrating the RG equation leads to such a resummation 
was, together with the self--energy equation, 
crucial for accessing the symmetry--broken regime 
with a fermionic flow in \cite{SHML}. The above structure 
suggests that this can also be done using the dynamic Wick 
ordered hierarchy developed here.  

In this example, one resums in the 
particle--particle channel, so  $\Pi (s)$ is simply 
\begin{equation}
\Pi_{U,V,U',V'} (s)
=
R(s)_{U,V} \; R(s)_{U',V'}
\end{equation}
where $R(s)$ denotes the non--charge--invariant propagator
with $A(s)$ given by the matrix in (37), (38) of \cite{SHML},
multiplied by the infrared cutoff function.
However, there is a crucial difference to the 1PI scheme,
which I now discuss at the example of the flow of superconductor
done in \cite{SHML}.  
By definition, the Wick ordering propagators  $R(s)$ are supported
at small scales below $\epsilon_0 \E^{-s}$, so 
$\lim_{s\to\infty} R(s) =0$. Thus $\lim_{s \to \infty} \Pi (s) =0$, 
and \Ref{wiso} gives for $s \to \infty$ 
\begin{equation}\label{achso}
W_\infty
=
W(0) \;
[1+ \Pi (0) W(0)]^{-1} .
\end{equation}
That is, the solution of the channel RG equation \Ref{wiso}
does not lead to the final, but the {\em initial} value of the 
propagator in the solution of the four--point function. 
Hence, in the solution of the BCS model given by this truncation,
it is the initial gap that appears in the solution for the four--point function.
This seeming  contradiction (when viewed from the perspective of \cite{SHML},
where the initial gap is put in as a symmetry--breaking parameter that is 
subsequently sent to zero) is resolved when one remembers that
in the Wick ordering scheme  one also has to Wick order the initial interaction. 
The requirement that the initial interaction is Wick ordered 
{\em but has no quadratic part in the fields} leads to the Hartree--Fock 
self--consistency equation. In the reduced BCS model, this equation yields the exact
solution for the superconducting gap. 
That is, in the Wick ordered dynRG, the initial Wick ordering of the 
bare interaction already gives the exact solution of the model, so the gap is 
constant and equal to the initial one. The flow 
of $\cW^{(2)}$ then gives the exact vertex. 
The two--loop contributions to $Q$, 
as well as the higher contributions to $\cW^{(2)}$, all vanish in the 
thermodynamic limit, as explained in the Appendix of \cite{SHML}.

\subsection{Relation to discrete RG flows}\label{sec:2.9}

In this section, I discuss a discrete analogue of the 
flow differential equation, and show that the differential
flow with a dynamically changing propagator can be obtained
as a limit of this discrete equation. 
The successive changes of the propagator in the discrete scheme
take the form of a continued fraction expansion for operators.

This approach gives a specific way to check the existence of the solution
to the equation \Ref{ivp2}, which is fundamental for the flow developed here. 
A full mathematical proof of existence is rather nontrivial, but the following
gives a clear prescription how to proceed and what to check. Discrete flows
are interesting by themselves since they are used in the mathematical 
studies in this field. 

\subsubsection{Adaptive iteration}
Consider again the effective action $\cG$ for a system of fermions with propagator
$D$ and interaction $\cV$, defined by
\begin{equation}
\E^{\cG(D_0,\cV_0,\Phi)}= \int \dd\mu_{D_0}(\Psi) \E^{\cV_0(\Psi+\Phi)} .
\end{equation}
$\cG$ is of the form
\begin{equation}\label{Gform}
\cG(F,V,\Psi) = \cK(F,V) + \frac12 (\cQ(F,V) \Psi, \Psi)
+ \Four (F,V)(\Psi) ,
\end{equation}
where $\cK$ is independent of $\Psi$,
$\cQ$ defines a quadratic form that gives the
quadratic term of $G$ in $\Psi$,
and $\Four$ contains all the higher powers of the fields.

By successively splitting the covariance in two pieces,
integrating over fluctuations, and shifting quadratic parts 
into the measure, I will now 
recursively construct a sequence of propagators and
interactions $(D_n,\cV_n)_{n \ge 1}$
with the property that the quadratic part of $\cV_n$ vanishes. 
Assume that $\cV_0$ has no constant and
quadratic parts. Split $D_0 = E_0 + F_0$,
then 
\begin{equation}\label{semigroup}
\E^{\cG(D_0,\cV_0,\Phi)} 
=
\int \dd\mu_{E_0} (\Psi_1) \;
\E^{\cG(F_0,\cV_0,\Psi+\Phi)} .
\end{equation}
By \Ref{Gform}, 
$\cG(F_0,\cV_0)$ splits into a 
constant term $K_0 = \cK(F_0,\cV_0)$, a
quadratic term $Q_0=\cQ(F_0,\cV_0)$, 
and a term $\cV_1=\Four (\cG(F_0,V_0))$.
The Gaussian measure is
$\dd\mu_{E_0}(\Psi) = \cN( E_0) \E^{-\frac12 (\Psi, E_0^{-1} \Psi)} D\Psi$, 
with the normalization factor $\cN$ for the Gaussian measure 
being a power of determinant or Pfaffian of $E_0$, 
so the quadratic term given by $Q_0$ can be shifted into the measure, 
and get
\begin{equation}
\dd\mu_{E_0}(\Psi)\; \E^{\frac12 (\Psi, Q_0 \Psi)}
 = 
\cN (1-E_0 Q_0) \; \dd\mu_{D_1} (\Psi)
\end{equation}
with the propagator
\begin{equation}\label{D1def}
{D_1}^{-1} = {E_0}^{-1} - Q_0 ,
\end{equation}
which now includes the corrections to the selfenergy given by $Q_0$. 
Note that $Q_0$ is not the proper selfenergy $\Sigma_0$ but instead 
the connected amputated two--point function. 
The relation to the selfenergy $\Sigma_0$ coming 
from the integration over $\Psi_0$ is 
\begin{equation}
Q_0 = (1-\Sigma_0 F_0)^{-1}\, \Sigma_0
\end{equation}
If one takes the limit of a strict support condition for $F_0$,
$Q_0=\Sigma_0$ outside the support of $F_0$, i.e.\ at 
low energy scales.

In writing \Ref{D1def}, I have assumed that the RHS is invertible.
This condition will be further discussed below. 
An obvious variant of this formula holds if one decides to put
only part of $Q_0$ (e.g.\ the part that shifts the Fermi surface)
into the propagator. 

In the convolution integral on the right hand side of \Ref{semigroup},
$\Phi+\Psi$ appears instead of $\Psi$, so that
\begin{equation}\label{eq:74}
\E^{-G(D_0,V_0,\Phi)} 
=
\E^{\tilde K + \frac12 (\Phi,Q_0\Phi)} \;
\int\dd\mu_{D_1} (\Psi) \;
\E^{(\Psi,Q_0\Phi) + {\cV_1} (\Psi + \Phi)} 
\end{equation}
with $\tilde K_0 = K + \log \cN (1 - E_0 Q_0)$.
Completing the square in the Gaussian measure and using the identity
$1+D_1Q_0 = D_1 {E_0}^{-1}$,
%and  $Q_0+Q_0D_1Q_0 = {E_0}^{-1} D_1 Q_0$, 
gives
\begin{equation}
\cG(E_0+F_0,\cV_0,\Phi ) 
=
\tilde K_0 + \frac12 (\Phi, (Q_0+Q_0D_1Q_0) \Phi)
+
\cG(D_1,\cV_1,D_1{E_0}^{-1} \Phi) .
\end{equation}
Iteration of this identity by splitting $D_1=E_1+F_1$ and
proceeding as above gives 
\begin{eqnarray}
\cG(D_0,\cV_0,\Phi) 
&=&
\sli_{l=0}^{n-1} \tilde K_l +
\frac12 \sli_{l=0}^{n-1} (S_l \Phi, Q_l (1+ D_{l+1}Q_l) S_l \Phi)
\nonumber \\
&+&
\cG(D_n,\cV_n,S_n\Phi)
\end{eqnarray}
with recursively determined propagators $D_n$,
interactions $\cV_n$ and amputation operators 
\begin{equation}
S_n = \pli_{k=0}^{n-1} D_{k+1} E_k^{-1}.
\end{equation}
Consequently,
\begin{equation}
\log P (D_0,V_0,\ETA) =
K_n + \frac12 (G_n \ETA, \ETA) +
G(D_n,V_n,S_n D_0 \ETA)
\end{equation}
with 
\begin{equation}\label{Cndef}
G_n = D_0 - \sli_{l=0}^{n-1} (S_lD_0)^T  Q_l (1+ D_{l+1}Q_l) S_lD_0 .
\end{equation}
If the splitting in $E_n$ and $F_n$ is chosen such that
$D_n \to 0$ as $n \to \infty$, and if $K_n$, $\cV_n$, $S_n$, and $G_n$ converge
to limits $K$, $\cV$, $S$, and $G$ in that limit, 
\begin{equation}
\log P (D_0,\cV_0,\ETA) = K 
+
\frac12 (G \ETA,  \ETA) 
+
\cV(S \ETA).
\end{equation}
If at each iteration step the full $Q_n$ was put into the measure
to define $D_{n+1}$, $G$ is the full propagator of the model,
and $\cV$ generates the connected $m$--point functions of the model
with $m \ge 6$, with amputation given by $S$. 
The interpretation at a finite $n$ is similar, except that 
there still remains an effective interaction for the degrees of freedom
that have not yet been integrated over.

In the following, I discuss two specific prescriptions 
for the splitting, which cover most of the applications,
and derive the corresponding formulas
for $S_n$, $D_n$, and $G_n$.

\subsubsection{Difference cutoff}\label{diffrgsect}
Assume that the original 
covariance $D_0$ is given in terms of an invertible $A_0$ 
as $D=A_0^{-1}$, and that $A_0$ commutes with its adjoint 
$A_0^*$. The operator $|A_0|^2=A_0^*A_0$ is hermitian and positive
and therefore has a spectrum that is a subset of 
$\bR_0^+=\{ x \in \bR: x \ge 0\}$. 
For $n \ge 1$, let $\gchi_n + \lchi_n =1$ be a $C^\infty$--partition of
unity on $\bR_0^+$, where $\lchi_n$ is decreasing and 
$\gchi_n$ is increasing and both are strictly positive 
functions. We also require that 
\begin{equation}
\frac{\lchi_{n+1}(x)}{\lchi_{n}(x)} \le 1 ,
\end{equation}
the idea being that the $\lchi_n$ provide lower and
lower cutoffs as $n$ increases. A possible choice is
$\lchi_n (x) = (1+\E^{\gamma_n (x-1)})^{-1}$ and
$\gchi_n=1-\lchi_n$, but the details of this choice 
do not matter for the moment.
The operator $\lchi_n (A^* A) $, defined via the spectral
representation of $|A|^2=A^*A$, is a 
positive operator that cuts off the parts of the spectrum
outside the support of $\lchi_n$. It is not a projection
because we chose a continuous partition of unity
instead of a step function. With this choice of $\lchi$
as pointwise strictly positive, $\lchi_n (|A|^2) $
is even invertible. 

Given $A_0$ and the interaction $\cV_0$, 
the sequence is constructed as follows.  
Set 
\begin{equation}
E_0 = {A_0}^{-1}\; \lchi_1(|A_0|^2).
\end{equation}
Then $F_0 = D_0-E_0 = {A_0}^{-1} \gchi_1(|A_0|^2)$. 
Calculate $Q_0 = \cQ(F_0,\cV_0)$ and $\cV_1=\Four (F_0,\cV_0)$,
and define
\begin{equation}
A_1 = A_0 - \lchi_1(|A_0|^2)\, Q_0.
\end{equation}
If $A_1$ is invertible, set $D_1 = {A_1}^{-1}\, \lchi_1 (|A_0|^2)$.
Then $D_1^{-1}=E_0^{-1}-Q_0$. 
In general, for $n \ge 1$, and given $\cV_n$ and an invertible $A_n$, 
set
\begin{equation}
D_n = {A_n}^{-1}\; \lchi_n(|A_{n-1}|^2)
, \quad
E_n = {A_n}^{-1}\; \lchi_{n+1}(|A_n|^2),
\end{equation}
and $F_n=D_n-E_n$. Determine $Q_n = \cQ(F_n,\cV_n)$ and
$\cV_{n+1} = \Four (F_n,\cV_n)$, and set 
\begin{equation}
A_{n+1} = A_n - \lchi_{n+1}(|A_n|^2) \, Q_n.
\end{equation}
If $A_{n+1}$ is invertible, the iteration can be continued;
otherwise, the iteration stops at $n$. 
The thus defined $D_n$ and $E_n$ satisfy
\begin{equation}\label{DEQ}
{D_n}^{-1} = E_{n-1}^{-1} - Q_{n-1}
,\quad
D_{n+1} E_n^{-1} = A_{n+1}^{-1} A_n,
\end{equation}
%and 
%\begin{equation}\label{tel}
%D_{n+1} E_n^{-1} = A_{n+1}^{-1} A_n,
%\end{equation}
so that 
\begin{equation}
\prod\limits_{n=0}^{N-1} D_{n+1} E_n^{-1} = A_{N}^{-1} A_0.
\end{equation}
The condition that $A_n$ is invertible has to be checked 
in every step because it is, in general, nontrivial,
and its failure can mean that there is some instability.
We discuss this in more detail in the application 
to the Fermi surface deformation below. 

There is another, less obvious, property of this scheme:
in the fluctuation propagator  
\begin{equation}\label{Fdef}
F_n =A_n^{-1}\; 
\Big(\lchi_n(|A_{n-1}|^2)
- \lchi_{n+1}(|A_{n}|^2)\Big) ,
\end{equation}
the difference of cutoff functions does not necessarily
give a positive operator. In a bosonic theory, this would
already make the fluctuation integral ill--defined.
For fermions, lack of positivity is not a problem for the definition
of the fluctuation integral, 
but numerically,  positivity is important. 
Roughly speaking, the difference of cutoff functions becomes
negative if the change of the propagator is so strong that
one starts integrating backwards. 

One can modify the cutoff function by replacing $A$ by $PA$,
where $P$ is some projection, whenever it appears as the 
argument of a cutoff function. 
For instance, in the fermion models, one can keep the 
cutoff function independent of the Matsubara frequency variables.

\subsubsection{Positive cutoff function}
The following alternative scheme is more straightforward
but does not admit a continuum limit if one wants to keep
a continuous partition of unity. 

Given a  single partition of unity $\gchi + \lchi =1$,
with $\lchi$ decreasing, 
and a strictly decreasing sequence
of positive numbers $(\Escale_n)_{n \ge 1}$, 
let $\lchi_1 = \lchi \left( \frac{|A_0|^2}{\Escale_1^2}\right)$
and $\gchi_1 = 1-\lchi_1$.
Set 
\begin{equation}
E_0 = A_0^{-1} \lchi_1, \qquad F_0 = A_0^{-1} \gchi_1.
\end{equation}
Then $E_0+F_0=D_0$, and both have positive cutoff operators. 
Again, calculate $Q_0 = \cQ(F_0,\cV_0)$ and $\cV_1=\Four (F_0,\cV_0)$,
and set $A_1 = A_0 - \lchi_1\, Q_0$. 
If $A_1$ is invertible, set $D_1 = A_1^{-1}$.

We now proceed in exactly the same way, by splitting
$D_1 = D_1 (\lchi+\gchi) = E_1+F_1$ and so on. 
This builds up a product of cutoff functions, 
which remains positive because each factor is positive. 
More precisely, for $n \ge 1$, and given $A_0, \ldots, A_{n}$, let
\begin{equation}
\lchi_n = 
\lchi_{n-1} \,
\lchi
\left( \frac{|A_{n-1}|^2}{\Escale_{n}^2}\right)
=
\prod\limits_{k=1}^n 
\lchi
\left( \frac{|A_{k-1}|^2}{\Escale_k^2}\right)
\end{equation}
and set 
\begin{equation}
D_n = A_n^{-1} \lchi_n ,
\qquad
E_n = A_n^{-1} \lchi_{n+1}.
\end{equation}
Then both $E_n$ and 
\begin{equation}
F_n = D_n - E_n =
A_n^{-1} 
\lchi_n
\; 
\gchi \left( \frac{|A_{n}|^2}{\Escale_{n+1}^2}\right)
\end{equation}
have a positive cutoff factor. 
After calculation of $Q_n$, set 
\begin{equation}
A_{n+1} = A_n - \lchi_{n+1} Q_n
\end{equation}
and, if $A_{n+1}$ is invertible (which is the case
if $Q_n$ is small enough), 
$D_{n+1} = A_{n+1}^{-1} \lchi_{n+1}$.
Again, this sequence satisfies \Ref{DEQ}. 

\subsubsection{The summed iteration}
Both of the above schemes can be described as follows.
We have a sequence of invertible operators $A_n$ and
positive operators $\lchi_n$, such that $A_0=A$, 
$\lchi_0=1$, and $D_n =A_n^{-1} \lchi_n$, 
$E_n= A_n^{-1} \lchi_{n+1}$, and 
$A_{n+1}= A_n -\lchi_{n+1} Q_n$, 
with $Q_n$ given by the fluctuation integral
with propagator $F_n=D_n-E_n$ and interaction $\cV_n$. 
Iteration gives 
\begin{eqnarray}
\cG(D_0,\cV_0,\Phi) &=&
\sum\limits_{l=0}^{n-1} \tilde K_l +
\frac12 \sli_{l=0}^{n-1} 
\left( 
A_l^{-1} A_0 \Phi, \;
(Q_l + Q_l D_{l+1} Q_l )
A_l^{-1} A_0 \Phi
\right)
\nonumber
\\
&+&
\cG(D_n,\cV_n,A_n^{-1} A_0 \Phi) .
\end{eqnarray}
This equation has an easy interpretation: the $\cG$ on the RHS
contains a fluctuation integral over fields with propagator
$D_n$ and interaction $\cV_n$, and the external field is 
amputated with $A_n$ instead of $A_0$, as it should be, 
since $D_n$ is $A_n^{-1}$ times a cutoff function. 
The constant and quadratic terms are the contribution to the
free energy density and the full propagator. 

The generating functional is thus given by 
\begin{equation}
\log P (D_0,\cV_0,\ETA) =
\bar K_n +\frac12 (G_n \ETA, \ETA) 
+ \cG(D_n,\cV_n,A_n^{-1}\ETA)
\end{equation}
where $\bar K_n$ is the field--independent term and
\begin{equation}
G_n =
D_0 -
\sli_{l=0}^{n-1} 
A_l^{-1} (Q_l + Q_l D_{l+1} Q_l) A_l^{-1}
\end{equation}
is the contribution to the full propagator
up to step $n$ in the RG iteration. 

\subsubsection{Limit of a differential equation}
I now use the "difference cutoff" scheme 
of Section \ref{diffrgsect}
to obtain a continuum limit of the RG sequence.
Let $\Escale_0 > 0$ be an initial energy scale, let
$
\Escale_n = \Escale_0 - n \veps
$
and choose 
$
\lchi_n(x) = \lchi(\frac{x}{\Escale_n^2})
$
where  $\lchi+\gchi=1$ is a fixed partition of unity on $\bR_0^+$
with strictly positive functions $\lchi$ and $\gchi$. 
Since there is a one--to--one relation between $n$ and $\Escale_n$
and we want to take a continuum limit, we now label the sequence
by $\Escale_n$ instead of $n$, and denote
$A_n = A_{\Escale_n}$, $\cV_n = \cV_{\Escale_n}$, etc.
Since in the limit $\veps \to 0$ nothing is integrated over,
$Q_{\Escale_n}$ is of order $\veps$:
\begin{equation}
Q_{\Escale_n} = \veps \, Q'_{\Escale_n}.
\end{equation}
Because $\Escale_{n\pm1} = \Escale_n \mp \veps$, 
the fluctuation propagator at $\Escale=\Escale_n$ is, 
by \Ref{Fdef}, 
\begin{equation}
F_{\Escale} = A_{\Escale}^{-1} 
\left(
\lchi \left(\frac{|A_{\Escale+\veps}|^2}{\Escale^2}
\right) 
-
\lchi
\left(\frac{|A_{\Escale}|^2}{(\Escale-\veps)^2}
\right)
\right)
\end{equation}
Thus, up to terms that vanish as $\veps\to 0$, 
$ F' _{\Escale} =  \sfrac{1}{\veps} F_{\Escale}$ is
given by 
\begin{equation}
 F' _{\Escale} 
=
A_{\Escale}^{-1} \; 
 \frac{\partial}{\partial\Escale}
\lchi'
\left(\sfrac{|A_{\Escale}|^2}{\Escale^2}
\right)
\end{equation}
Note that the $\Escale$--derivative acts only on the 
cutoff function, but not on the inverse in front.
%The derivative gives 
%\begin{eqnarray}
%F'_{\Escale} 
%&=&
%A_{\Escale}^{-1} \; (-2\lchi') 
%\left(\sfrac{|A_{\Escale}|^2}{\Escale^2}
%\right)
%\\
%&&
%\left\lbrack
%\frac{|A_{\Escale}|^2}{\Escale^3}
%- \frac{1}{2\Escale^2}
%\left(
%{Q'_{\Escale}}^* \lchi(\sfrac{|A_{\Escale}|^2}{\Escale^2})
%A_{\Escale} +
%A_{\Escale}^* \lchi(\sfrac{|A_{\Escale}|^2}{\Escale^2})
%Q'_{\Escale}
%\right)
%\right\rbrack .
%\end{eqnarray}
The differential equation for $Q_\Escale$ and the other functions
is obtained by doing the fluctuation integral to $O(\veps)$. 
Using \Ref{Lappi}, 
%The Gaussian convolution can be written in terms of the 
%functional Laplacian
%\begin{equation}
%\Delta_F = \frac12 (\frac{\delta}{\delta \Phi},
%\; F \frac{\delta}{\delta \Phi} )
%\end{equation}
%to get \cite{crg}
%\begin{equation}
%\E^{G(F_n,\cV_n,\Phi)} = \int \dd\mu_{F_n}(\Psi) \E^{\cV_n(\Psi+\Phi)}
%=
%\E^{\Delta_{F_n}} \E^{\cV_n(\Phi)} .
%\end{equation}
recalling that $F_n = F'_{\Escale_n} \veps$, and dropping 
terms of order $\veps^2$, I get 
\begin{eqnarray}
\cG(F_n,\cV_n,\Phi) 
&=&
\log \left( (1+ \Delta_{F_{\Escale_n}}) \E^{\cV_n(\Phi)}\right)
\nonumber\\
&=&
\log \left( \E^{\cV_n(\Phi)} 
(1+ \E^{-\cV_n(\Phi)}\Delta_{F_{\Escale_n}} \E^{\cV_n(\Phi)})\right)
\nonumber\\
&=&
\cV_n(\Phi) + 
\E^{-\cV_n(\Phi)}
\Delta_{F_{\Escale_n}}
\E^{\cV_n(\Phi)} .
\end{eqnarray}
Thus 
\begin{equation}
\cG(\veps F'_{\Escale},\cV_{\Escale},\Phi) 
=
\cV_\Escale(\Phi) + 
\veps
\left\lbrack
\Delta_{F'_\Escale} \cV_\Escale (\Phi) +
\frac12
\left(
\frac{\delta \cV_\Escale}{\delta\Phi},\;
F'_\Escale
\frac{\delta \cV_\Escale}{\delta\Phi}
\right)
\right\rbrack
\end{equation}
hence the increment is simply given by the right hand side of 
Polchinski's equation. The main difference here is, however, 
that $\cV_\Escale$ has no quadratic part.
Writing $\cV_\Escale = \cV_\Escale^{(4)} + \cV_\Escale^{(\ge 6)}$,
extracting the quadratic part of $\cG$, % as in \Ref{Gform},
and taking the limit $\veps \to 0$, gives the differential
flow equation as follows. The change of the inverse propagator is 
given by
\begin{equation}\label{Argde}
\sfrac{\partial}{\partial \Escale} 
A_\Escale 
= 
Q'_\Escale \lchi(\sfrac{|A_{\Escale}|^2}{\Escale^2})
\end{equation}
where $Q'_\Escale$ is determined by
\begin{equation}
\frac12 (\Phi, Q'_\Escale \Phi) 
=
\Delta_{F'_\Escale} \cV_\Escale^{(4)} (\Phi)
\end{equation}
The interaction part obeys
\begin{equation}\label{Vge6de}
\sfrac{\partial}{\partial \Escale} 
\cV_\Escale (\Phi) 
=
\Delta_{F'_\Escale} \cV_\Escale^{(\ge 6)} (\Phi)
+ 
\frac12
\left(
\delta_\Phi\cV_\Escale,\;
F'_\Escale
\delta_\Phi\cV_\Escale
\right)
\end{equation}
These are the equations obtained directly in the continuum in Section \ref{rgdesec}
(with the change of variables $\Escale \to s=\log \Escale_0/\Escale$).

\subsubsection{Wick ordering in the discrete scheme}
The discrete scheme also provides insights about the choice of 
Wick ordering in the limit of a differential equation, and it is 
of independent interest. The point where things start to 
differ from the non--Wick--ordered scheme is the definition of $D_1$.
Instead of  \Ref{D1def}, $D_1$ is now defined such that the integral
in \Ref{eq:74} is rewritten in terms of a Wick ordered interaction. 
%\begin{equation}
%\int \dd\mu_{(E_0^{-1} - Q_0)^{-1}} (\Psi) 
%\E^{(\Psi, H ) + \cV_1 (\Psi + \Phi)}
%=
%\E^{\tilde K_1}
%\int \dd\mu_{D_1} (\Psi)
%\E^{(\Psi, H ) + \Omega_{D_1} (\cW_1) (\Psi + \Phi)}
%\end{equation}
%
More precisely, first set $\Phi=0$ in \Ref{eq:74} and require that
\begin{equation}\label{eq:seWi1}
\int \dd\mu_{(E_0^{-1} - Q_0)^{-1}} (\Psi) 
\E^{\cV_1 (\Psi)}
=
\E^{\tilde K_1}
\int \dd\mu_{D_1} (\Psi)
\E^{\Omega_{D_1} (\cW_1) (\Psi)}
\end{equation}
where $\Omega_D (F) = \E^{-\half \Delta_D} F$ denotes
Wick ordering of $F$, as in \cite{crg}, and 
\begin{equation}
\cP^{(\le 2)} \cW_1 = 0,
\end{equation}
i.e.\ $\cW_1(\Psi)$ does not have constant or quadratic parts in $\Psi$.
This last condition leads to a self--consistency equation for $D_1$
whose solution differs from \Ref{D1def}. By definition of the Gaussian measure, 
this means that
\begin{eqnarray}
&&
\log \cN( (E_0^{-1} - Q_0)^{-1})
-
\half
( \,(E_0^{-1} - Q_0)\Psi, \; \Psi)
+
 \cV_1 (\Psi) 
\nonumber\\
&=&
\log \cN (D_1) 
+ 
\tilde K_1
-
\half
(D_1^{-1} \Psi, \;  \Psi)
+
 \E^{-\half \Delta_{D_1}} \cW_1 (\Psi) 
\end{eqnarray}
$\cW_1$ is obtained by using $1 = \cP^{(\le 2)} + \cP^{(\ge 4)}$ to rewrite
\begin{equation}\label{eq:1erlei}
\cV_1 
=
 \E^{-\half \Delta_{D_1}}\; 
\left(
\cP^{(\le 2)} + \cP^{(\ge 4)}
\right)
\;
\E^{\half \Delta_{D_1}}\; \cV_1 .
\end{equation}
Since $\cP^{(\ge 4)}$ projects out all terms constant or quadratic in the fields, 
this gives 
\begin{eqnarray}
\cW_1
&=&
\cP^{(\ge 4)} \; \E^{\half \Delta_{D_1}}\; \cV_1
\nonumber\\
\tilde K_1
&=&
\cP^{(0)}
\E^{\half \Delta_{D_1}}\; \cV_1
-
\half \Delta_{D_1} \cP^{(2)} \E^{\half \Delta_{D_1}}\; \cV_1
\end{eqnarray}
and 
\begin{equation}\label{eq:110}
D_1^{-1}
=
E_0^{-1} - Q_0  +
\bP^{(2)}
\E^{\half \Delta_{D_1}}\; \cV_1
\end{equation}
where for a function of the fields $F(\phi)$, 
$A=\bP^{(2)} F$is the operator in the quadratic form 
given by $\cP^{(2)} F$, i.e.\ $\cP^{(2)} F (\Phi)  = \half ( A\; \Phi, \Phi)$. 
Equation \Ref{eq:110} is the self--consistency equation for $D_1$. 
It takes the form of a generalized Hartree--Fock--type equation 
because the action of  $\E^{\half \Delta_{D_1}}$ is to create self--contractions 
of the vertex given by $\cV_1$ with a $D_1$ propagator, and this 
feeds into $D_1^{-1}$ in the way that is standard in self--consistency
equations of mean--field type. 
Now that $\cW_1$ and $D_1$ are given,  
one obtains the rewriting of the integral in 
\Ref{eq:74} for $\Phi \ne 0$ 
by standard shifting of Gaussians and redefinitions of 
the amputation term that acts on the external field $\Phi$. 

\section{Fermi Surface Flows}\label{fsfsec}
In this section, I briefly discuss the Fermi surface flow in $d \ge 2$ in the scheme 
without Wick ordering. A discrete Fermi surface flow with Wick ordering 
is defined and controlled mathematically for $d=2$ in \cite{PeSa}.
This is one of the examples that motivated the method. 
Although different methods, such as the $T$--flow of \cite{Tflow}
have been used to calculate RG flows,  and other regularizations, 
such as a cutoff on the frequencies only, are possible, 
the momentum space flow remains the only one where 
precise mathematical estimates for power counting of all Green functions, 
not just the first few truncations, have been given \cite{FST}. 

Let the covariance and interaction be those of the standard
many--fermion system with short--range interaction $\cV_0$,
as described in the Introduction, with $n \to \infty$. 
(for more details, see \cite{msbook}, Chapter 4). 
In Fourier space, the covariance is a multiplication operator, 
namely the inverse of 
\begin{equation}
A_0 (q_0,q) = \I q_0 - e_0 (q) ,
\end{equation}
times the unit matrix in spin space. 
Here $e_0$ is the dispersion relation for free particles, 
given by the hopping amplitudes of the model. 
By standard (formal) diagrammatics, the full propagator is the 
inverse of $\I q_0 - e_0 (q) - \sigma(q_0,q)$, with $\sigma$ 
the fermion self--energy. 
One natural choice for would then be to take $A$ as the full 
propagator, with $\sigma$ understood to depend on $s$
(but without a cutoff function). 
In the following, I shall discuss another choice, where only 
the frequency zero part of the selfenergy is put into the 
denominator. This then implies that $V$ still has a quadratic part, 
but one which vanishes at zero frequency. In cases where the 
field renormalization is a marginally relevant parameter, 
one can also put in the linear part of $\sigma $ in $q_0$. 
Thus the freedom in chosing $Q$ and $A$ allows to adapt to 
the situation at hand, while keeping the denominator as 
simple as possible.  

So, let 
\begin{equation}
A(q_0,q) 
=
\I q_0 - e(s,q) .
\end{equation}
In terms of the scale--dependent self--energy $\si_s (q_0,q)$,
$e(s,q) = e_0(q) + \sigma_s(0,q)$, so that the difference
\begin{equation}
\tilde \sigma_s (q_0,q) = \sigma_s (q_0,q)  - \sigma_s (0,q) 
\end{equation}
vanishes at $q_0 = 0$. Let $\tilde \chi$ 
be a decreasing cutoff function with $\tilde \chi(x) = 1$ 
for $x \le 1/2$ and $\tilde \chi(x) = 0$ for $x \ge 1$. 
Set 
\begin{equation}\label{ada}
\chi(s,k) 
=
\tilde\chi 
\left(
\frac{e(s,k)^2}{\Esc{s}^2}
\right) .
\end{equation}
(another choice would be to use a cutoff on the frequencies as well:
$
\chi(s,k_0,k) 
=
\tilde\chi 
\left(
\frac{k_0^2+e(s,k)^2}{\Esc{s}^2} 
\right) 
$.

With this choice, $\cV(s)$ still has a quadratic part, which is of the form
\begin{equation}
\int \db \dpo{k} \; 
\bar\psi (\dpo{k}) v_1 (s,\dpo{k} ) \psi (\dpo{k})
\end{equation}
with 
\begin{equation}\label{nullfi}
v_1 (s, (0,k)) = 0
\end{equation}
(here 
$\dpo{k} = (k_0,k)$, and $\int \db \dpo{k} = $
$\beta^{-1} \sum_{k_0 \in M_f}\int \frac{\dd k}{(2\pi)^d} $
with $M_f $ the set of fermionic Matsubara frequencies).
Thus with this choice of moving only frequency--inde\-pen\-dent 
parts of the selfenergy into the denominator, there remain 
some quadratic terms in the action. But by construction, 
these terms vanish on the Fermi surface (see \Ref{nullfi}), 
and hence do not cause divergences.

The choice \Ref{ada} leads to an adaptive scale decomposition
because $\chi$ is supported in the neighbourhood 
\begin{equation}
\{ k : \abs{e(s,k)} \le \Esc{s} \} 
\end{equation}
of the set $\{ k :  e(s,k) = e_0(k) + \sigma_s (0,k) = 0 \}$
which one may regard as a flowing Fermi surface. 
In the standard, nonadaptive scale decomposition, 
the cutoff function is taken as 
$\chi_0 (s,k) = \tilde\chi(e_0(p)^2\Esc{s}^{-2})$, 
whose support shrinks to the free Fermi surface as $s \to \infty$. 
Consequently, the nonadaptive scale decomposition can be used only
if a counterterm is included \cite{FT,FST} -- or if the Fermi surface
shift is ignored, as is sometimes the case in approximate studies.

In the present case, the terms in \Ref{homexp} are
\begin{eqnarray}
V^{(m)} (\Psi)
&=&
\int \prod\limits_{n=1}^{m} \db \dpo{k}_n \, \db \dpo{k}'_n
\bar\psi (\dpo{k}_n) \psi (\dpo{k}'_n) \quad
\delta (\sum (\dpo{k}_n - \dpo{k}'_n)) \;
%\prod\limits_{n=1}^m \bar\psi (\dpo{k}_n) \psi (\dpo{k}'_n)
\nonumber\\
&&
V_m(s\mid \dpo{k}_1, \ldots, \dpo{k}_m ; \dpo{k}'_1, \ldots, \dpo{k}'_{m-1})
\end{eqnarray}
Equation \Ref{ivp2} can now be rewritten as 
\begin{equation}\label{dote}
\dot e (s,p) 
=
Q(s,p) \chi (s,p)
\end{equation}
with 
\begin{equation}
Q(s,p)
=
\int \db \dpo{k} \; 
F(\dpo{k})\;
V_2 (s \mid (0,p), \dpo{k}; \dpo{k})
\end{equation}
The fluctuation propagator 
\begin{eqnarray}\label{eq:122}
F (\dpo{k})
&=&
A(\dpo{k})^{-1} 
\frac{\del}{\del s} \chi (s,k)
\nonumber\\
&=&
A(\dpo{k})^{-1} 
\tilde \chi' 
\left(
\frac{e(s,k)^2}{\Esc{s}^2}
\right)
\frac{2}{\Esc{s}^2}
(e(s,k)^2 + e(s,k) \dot e(s,k) )
\end{eqnarray}
contains a term proportional to $\dot e$, because of the 
adaptive nature of the scale decomposition given by 
this choice of $\chi$. 

To discuss the mathematical properties of the resulting equations, 
it is useful to introduce the ``tadpole'' operator. For a function 
$\Phi (p; \dpo{k})$ it is the linear operator 
\begin{equation}\label{Tad}
(\Tad_\Phi f) (p)
=
\frac{2}{\Esc{s}^2}
\int
\db \dpo{q}\; 
\Phi (p; \dpo{q})
\tilde \chi' 
\left(
\frac{e(s,k)^2}{\Esc{s}^2}
\right)
\frac{1}{\I q_0 - e(s,q)}
\;
f(q)
\end{equation}
acting on functions of $q$. With 
\begin{equation}
\Phi_V (p, \dpo{k}) 
=
\cV(s \mid (0,p), \dpo{k};\dpo{k})
\end{equation}
and 
\begin{equation}
\Phi_{V,e} (p, \dpo{k}) 
=
\cV(s \mid (0,p), \dpo{k};\dpo{k}) \; e(k)
\end{equation}
\Ref{dote} becomes
\begin{equation}\label{dote2}
\left(
1 - \Tad_{\Phi_{V,e}}
\right) (\dot e)
=
\Tad_{\Phi_{V}} (e^2)
.
\end{equation}
The operator $\Tad$ depends nonlinearly on $e$, but it is linear 
in $\dot e$. 

Thus already the differential equation for the change of $e$
involves an operator inversion ,  due to the additional 
term in the fluctuation propagator in \Ref{eq:122}, 
hence coming from the adapting scale decomposition. 
The solution strategby involves the following steps. 

\begin{enumerate}

\item
Show that $1-\Tad_{\Phi_{V,e}}$ has a bounded inverse
on some open set of $V$ and $e$.
This is essentially a restriction on the size of $V_2$ --- 
it will fail if $V_2$ gets so large that its backreaction 
on $e$ becomes of order 1. A sufficent condition is 
$\norm{\Tad} < 1$ in a suitable norm. This allows to rewrite 
\Ref{dote2} as 
\begin{equation}
\dot e = F(e) = 
\left(
1 - \Tad_{\Phi_{V,e}}
\right)^{-1}
\Tad_{\Phi_{V}} (e^2) .
\end{equation}
Because $\Tad$ depends on $e$, it is necessary to find a set of $e$
on which $\norm{\Tad} \le \eta < 1$ uniformly. 
The existence of this set already requires conditions on $e_0$;
the set is then essentially a ball around $e_0$. 

\item
The general theory of ordinary differential equations then implies existence 
and uniqueness of the solution for all $s$ in some bounded interval, 
provided that $F(e)$ is Lipschitz. Proving that this condition is
satisfied is again nontrivial, in fact it is closely related to 
questions of possible symmetry breaking. 

\item
To get global existence (i.e.\ for all $s$, one needs to prove that
the flow actually never leaves the ball around $e_0$. 

\item
In steps 1--3, $V$ was still regarded as fixed; in the true system of equations, 
$V_2$ depends on $s$ and the RHS of the equation for $\dot V_2$ 
depends on $e$ as well. Thus one needs to show that $V_2$ stays 
bounded. 

\end{enumerate} 
There is a similar set of steps for the discrete Fermi surface flow. 
Steps 1--3 are already nontrivial, but the hard part is including 4, because it 
requires dealing with the hierarchy as a whole. 
For the differential equation, an all--order proof that steps 1--4 work 
can be given along the lines of \cite{crg}. 
A nonperturbative proof, i.e.\ without truncation on the hierarchy, 
remains an open problem, 
mainly because to this day there are no good nonperturbative bounds
for the continous RG equation \cite{BWerratum}, which would 
be needed in step 4. Up to now, this step  
requires an integration over short scale intervals 
and corresponding estimates on discrete RG transformations \cite{SalWiecz}.
Such a proof is given in the Wick ordered approach, for two--dimensional
Fermi systems, in \cite{PeSa}. 
In particular, the boundedness of $V_2$ necessary in step 4
can be shown in the  temperature range of the Fermi liquid criterion of \cite{crg}. 

These mathematical points also correspond to physically natural problems. 
The question of the existence of the inverse of $1-\Tad$
is, in physical terms, basic to the consistency of the method itself:
the RG flow is parametrized by energy scales on the (changing)
kinetic part of the action, i.e.\ the part quadratic in the fields. 
The nonquadratic interaction part is not taken into account 
when labelling the scales. However, it influences the change $\dot e$
of the quadratic part, and the equations become inconsistent 
when this change becomes larger than the maximal kinetic 
energy scale itself. Note that it is this condition that is important in the flow;
the typical interaction energy, measured by the coupling constant, 
can be much larger than the scale $\Esc{s}$, and this situation 
indeed occurs at low scales in all nontrivial models because $\Esc{s} \to 0$
for $s \to \infty$. It is a nontrivial fact of fermionic field theory that 
the method can be controlled by convergent expansions even in that
situation \cite{SalWiecz}. 
The invertibility of $A(s)$ at all scales $s$ is related to this; the 
$s$--dependent Fermi surface must never leave the region where 
$\chi_< (\Esc{s}^{-2} e(s,k)^2 )\ne 0$, otherwise an inconsistency manifests itself as 
divergences in the equation.  This condition can be satisfied because
power counting implies that the selfenergy at scale $s$ is of order $\Esc{s}$
times the coupling constant (in fact, improved power counting \cite{FST,crg}
gives a better 
$s$--behaviour of the bound, which is needed to control derivatvies of 
the selfenergy in the flow). 

The counterterms used in \cite{FST} can be obtained from the adaptive flow:
if $\tilde e (k) = \lim_{s\to\infty} e(s,k)$ exists, the difference $\tilde e - e_0$ 
can be taken as the counterterm $K$ of \cite{FST}. This can be used to 
avoid solving the inversion equation \cite{PeSa}. However, the question whether there 
is a one--to--one relation between the free and the interacting dispersion function
and Fermi surface, which is answered by the inversion theorem,  
is not solved this way. 

The need to invert $1 - \Tad$ for a tadpole operator $\Tad$ 
also arises in the 1PI scheme when one reexpresses the 
single--scale propagator $S$ in terms of the full propagator $\dot G$
in the equation for the self--energy  $\Sigma$ \cite{Katanin,SHML}.
With the tadpole operator defined in a natural way, 
the equation for the selfenergy in the 1PI hierarchy becomes
\begin{equation}
\dot \Sigma
=
\Tad_{\Gamma^{(2)}} (S) 
=
\Tad_{\Gamma^{(2)}}
(\dot G - G \dot \Sigma G)
\end{equation}
where $\Ga^{(2)}$ is the 1-irreducible two--particle vertex and $G$ is the full
Green function.
This equation can be rewritten as 
\begin{equation}
(1- \Tad_{\Gamma^{(4)}GG}) (\dot \Sigma) 
=
\Tad_{\Gamma^{(4)}} (\dot G) .
\end{equation}
More generally, the occurrence of such operators can be expected in 
any scheme where the propagator changes, be it with an adaptive 
scale decomposition, as done here, or without (as in the usual 1PI hierarchy).

The flow discussed here can be adapted to a flow where the 
density, not the chemical potential $\mu$, is kept fixed. 
This variant was used in the Appendix of \cite{HSFR} to calculate the flow of the 
Fermi surface at fixed density in the Hubbard model. 

\section{Strong coupling behaviour in a toy model} 
\label{toysec}
In this section, I study the flow equations
for a very simple example, namely 
the integral over one complex variable,
corresponding to the partition function of 
a zero--dimensional complex bosonic field,
\begin{equation}\label{nulli}
P_0(h) 
=
 \sfrac{\az}{2\pi \I} 
\int_{\bC} \dd \bar\vphi \wedge \dd \vphi \;
\E^{-\az |\vphi|^2 - \laz |\vphi|^4 + h \bar \vphi + \bar h \vphi}
\end{equation}
with $\az > 0$, Re $\laz  > 0$ (for simplicity, 
I take $\laz $ real in the calculations below.), 
and $h$ a complex source field. 
For $\vphi=u+\I v$, $\dd \bar\vphi \wedge \dd \vphi = 2 \I \dd u \dd v$.
Thus it is simply a two--dimensional integral. 

This analysis is motivated by \cite{Meden}, where it was observed
that the 1PI RG equations give an accurate solution for a function 
similar to  $P_0(h)$ 
even for large $\la_0$. This is unexpected at first sight because 
the diagrammatic approach to these equations may suggest that 
they are useful only for small $\la_0$. 
Moreover, it is interesting because only the 
combination of RG and irreducibility seemed to induce the good
behaviour at large couplings, namely
(1) in contrast to the 1PI scheme, the other schemes studied in \cite{Meden},
namely Polchinski's original hierarchy and the Wick ordered hierarchy 
of \cite{crg}, were accurate only at small $\la_0$, 
as one would expect from small--coupling schemes
and (2) perturbation theory for the 1PI vertices $\Ga_m$ fails 
as badly as in the other schemes when the coupling constant is not small.
Thus it is not merely irreducibility, 
but a different feature of the 1PI RG hierarchy, 
that makes for the difference. 

A closer look at the equations reveals a first crucial feature: 
the 1PI equation contains the selfenergy $\si$ in the denominator.
When $\laz$ gets large, $\si$ becomes of order $\laz^{1/2}$ after a very 
short flow time already, and (because the signs conspire well
in this toy model) the factor $\laz$ in the numerator is balanced by a factor $\laz$
in the denominator, which prevents the right hand side of the flow equation
from getting very large. 
This suggests that a dynamic adjustment of the propagator, 
such as discussed in this paper, improves the behaviour of the other 
hierarchies as well. This is the case, but the above explanation
is not the full story, as we shall see in the following. 

\subsection{Asymptotics of $P$ at large $\la_0$}
Although we have written
$P_0$ as a function of $h$, it is understood that it also depends on 
$\bar h$. Indeed, by the $U(1)$ symmetry of the measure it depends 
only on $\bar h h =|h|^2$. 
A Gaussian transformation gives the representation
\begin{equation}\label{hubbstra}
P_0(h) 
=
\az \int _\bR\frac{\dd r}{\sqrt{4\pi\laz}}\;
\E^{-r^2/4\laz} \; 
\frac{1}{\az - \I r} \;
\E^{\frac{|h|^2}{\az-\I r}} 
\end{equation}
from which analyticity in $|h|$ is obvious. Write 
\begin{equation}
P_0 (h)  = 
\sum_{m \ge 0} \pi_m (\az,\laz) |h|^{2m} .
\end{equation}
Rescaling the integration variable in \Ref{nulli}, one gets
$
\pi_m (\az,\laz) 
=
\laz^{-\frac{m}{2}}\; 
%\frac{\az}{\sqrt{4\laz}} \; 
F_m \left( \frac{\az}{\sqrt{\laz}}\right)
$
where $F_m (\xi) $ is analytic in $\xi$ for $|\xi | < 1$. Explicitly, 
\begin{equation}
F_m(\xi)
=
\frac{1}{2(m!)^2} \; \xi 
\sum_{r=0}^\infty
\frac{(-1)^r}{r!}\;
\Ga \left( \frac{m+r+1}{2}\right) \;
\xi^r \; .
\end{equation}
This convergent expansion gives the behaviour at large $\laz$ as
\begin{equation}\label{eq:134}
P_0 (0) = \pi_0(\az,\laz)  
=
\frac{\sqrt{\pi}}{2}\; \frac{\az}{\sqrt{\laz}} + O (\frac{1}{\laz})
%\E^{w_0} = \sqrt{\frac{\pi \az^2}{4 \laz}} + O (\frac{1}{\laz})
\end{equation}
and 
\begin{equation}
\frac{P_0 (h)}{P_0(0)} 
=
\E^{ w_1 |h|^2 + w_2 |h|^4 + O (|h|^6)}
\end{equation}
with 
\begin{equation}\label{eq:136}
w_1 
=
\frac{1}{\sqrt{\pi}} \laz^{-1/2} + O (\frac{1}{\laz}), 
\qquad
w_2 
=
-
\left(
\frac{1}{2\pi} - \frac18
\right)
\;
\laz^{-1}
+ O(\laz^{-3/2}) .
\end{equation}
Thus for large $\laz$, the $2m^{\rm th}$ moments of the integral, 
which play the role of the $2m$--point functions in this toy model, 
become small. In the following I discuss to what extent this can be 
reproduced by the RG with propagator adjustment.

\subsection{The dynamical RG hierarchy for the toy model}
The usual shift gives
\begin{equation}
P_0(h) 
=
\E^{|h|^2/\az} \left( \mu_{\az^{-1}} * \E^{-w} \right)
\left(\frac{h}{\az}\right)
\end{equation}
with the notations
\begin{equation}
\dd \mu (\vphi) = \frac{\dd \bar\vphi \wedge \dd \vphi}{2\pi \I} \; \az \;
\E^{-\az |\vphi|^2}
, \quad
(\mu * F)(\phi) = \int \dd \mu(\vphi) F(\phi-\vphi)
\end{equation}
and $w (\vphi) = \laz |\vphi|^4 $.
As in the general case, write 
\begin{equation}
P(s,h) 
=
\E^{k(s) + g(s) |h|^2}
\left(
\mu_{d(s)} * \E^{- v(s)}
\right)
(\si(s) h)
\end{equation}
and impose 
$k(0)=0$, $g(0) = \az^{-1}$, $d_0 = \az^{-1}$, $\si(0)=\az^{-1}$, 
$v(0,\vphi) = w(\vphi)$, 
as well as $\del P/\del s = 0$. The RG equations are derived using
straightforward differentiation, in analogy to the general case. 
Expand 
\begin{equation}
v(s,\vphi)
=
\sum_{m \ge 0}
v_m (s) 
|\vphi|^{2m}
\end{equation}
then the equations read
\begin{equation}
\dot v_0
=
\dot k - d q - f v_1
,\quad
\dot v_1
=
q + f (v_1^2 - 4 v_2)
\end{equation}
and for $m \ge 2$
\begin{equation}
\dot v_m
=
f
\left(
\sum_{\mu=1}^m \mu (m+1-\mu) v_\mu v_{m+1-\mu} - (m+1)^2 v_{m+1}
\right)
\end{equation}
%\begin{eqnarray}
%\dot v_0
%&=&
%\dot k - d q - f v_1
%\nonumber\\
%\dot v_1
%&=&
%q + f (v_1^2 - 4 v_2)
%\nonumber\\
%\dot v_2
%&=&
%f (4 v_1 v_2 - 9 v_3)
%\nonumber\\
%\dot v_3
%&=&
%f (6 v_1v_3 + 4 v_2^2 - 16 v_4)
%\nonumber\\
%\dot v_m
%&=&
%f
%\left(
%\sum_{\mu=1}^m \mu (m+1-\mu) v_\mu v_{m+1-\mu}
%%\sum_{m_1,m_2 \atop m_1+m_2 = m+1}
%%m_1\, m_2\; v_{m_1}\, v_{m_2} 
%- (m+1)^2 v_{m+1}
%\right)
%\end{eqnarray}
The initial interaction $w$ has no constant and quadratic part. 
We now impose that the function $v$ has the same properties.
by requiring $\dot v_0 = 0$ and $\dot v_1 = 0$.
This gives equations for $q$ and $k$:
\begin{equation}
q = 4 f v_2 \quad \mbox{ and } \quad \dot k = dq .
\end{equation}
%The equations for larger $m$ now read 
%\begin{equation}
%\dot v_2 = - 9 f v_3, \qquad 
%\dot v_3 = 4 f v_2^2 - 16 f v_4,\qquad 
%\ldots
%\end{equation}
Finally, we fix the flow by setting 
\begin{equation}\label{deq}
d(s) = a(s)^{-1} \chi(s)
\end{equation}
where $a(s) > 0$ is to be determined and $\chi(s) > 0$ a
given function of $s$ that decreases from $\chi (0) = 1$ to 
$\chi(1)=0$ (here we take $s=1$ as the final value of the flow parameter.
In this simple example, we do not need to make 
$\chi$ depend on  $a$). We require $a$ and $\chi$ to be nonnegative to have
the same for $d$, so that the Gaussian measure is really a measure. 
Now set 
\begin{equation}\label{dotaeq}
\dot a = - \chi q
\end{equation}
Then $f=a^{-1} \dot \chi$. Note that $f$ is negative because 
$\chi $ is decreasing; the sign of $f$ is, however, unimportant 
because it never appears in a measure, only in a Laplacian.
With our initial condition, $v_2$ starts out positive; $v_2$ decreases
and $a$ increases in the flow. 

The system of equations reads 
\begin{eqnarray}
\dot k 
&=&
a^{-2} \chi\dot\chi 4 v_2
\label{dottker}\\
\dot a
&=&
-\chi\dot\chi\; 4 a^{-1} \; v_2
\label{dotter}\\
\dot v_2
&=&
- 9 a^{-1} \dot \chi v_3
\label{gelbesei}\\
\dot v_3
&=&
 a^{-1} \dot \chi
\left( 
4 v_2^2 - 16  v_4 
\right)
\label{dickesei}\\
\dot v_4 
&=&
a^{-1} \dot \chi
\left(
12 v_2 v_3  - 25 v_5
\right)
\label{grueneneune}\\
\dot v_m 
&=&
a^{-1} \dot \chi 
\left(
\sum_{\mu=2}^{m-1} \mu (m+1-\mu) v_\mu v_{m+1-\mu}
- (m+1)^2 v_{m+1}
\right)
\label{vmdot}
\end{eqnarray}
The summation in \Ref{vmdot} now excludes $\mu =1$ and $\mu = m$ because 
$v_1=0$. 
The equation for $a$ can be rewritten as 
\begin{equation}\label{aconvenient}
\frac{\del}{\del s}
(a(s)^2)
=
- 4 v_2(s) \; 
\frac{\del}{\del s}
(\chi(s)^2) .
\end{equation}
The ``amputation'' factor $\si (s)$ is now determined by the analogue of 
condition \Ref{eq:dodo}: putting  $\dot \si \si^{-1} - dq =0$ and inserting  
\Ref{deq} and \Ref{dotaeq}
gives $\si^{-1} \dot \si = - a^{-1} \dot a$, thus, by the 
initial condition, $\si(s) = a(s)^{-1}$.
Moreover, a glance at \Ref{dottker} and \Ref{dotter} reveals
that $\dot k = - a^{-1} \dot a$. Thus, once $a(s)$ is determined, 
\begin{equation}
k (s) = - \log \frac{a(s)}{\az} 
\quad\mbox{ and }\quad
 \si(s) = a(s)^{-1} 
\end{equation}
in accordance with the general results of Section \ref{diagrammar}.

\subsection{Truncations of the hierarchy}
There are still infinitely many equations for the $v_m$. 
We now study the accuracy of various truncations as a function of
$v_2 (0) = \la$. 
Suppose we truncate by setting $v_3 =0$. 
Then $v_2 (s) = \laz$ is a constant, and \Ref{aconvenient}  
integrates to 
\begin{equation}
a(s)= \sqrt{ \az^2 + (\chi(0)^2 - \chi(s)^2) 4 \laz} .
\end{equation}
Similarly, the equations for $k$ and $g$, which in this truncation read
\begin{equation}
\dot k 
=
\frac{4 \la \; \frac12 (\chi(s)^2)^\cdot}{\az^2 + (\chi(0)^2 - \chi(s)^2) 4 \la}
\quad\mbox{ and }\quad
\dot g
=
\frac{4 \laz \dot \chi}{(\az^2+ (1-\chi(s)^2) 4 \laz)^{3/2}} ,
\end{equation}
can easily be solved.
Because $\chi(0)=1$ and $\chi(1) = 0$, the final values 
$a_1=a(1)$, $k_1=k(1)$ and $g_1=g(1)$ are
\begin{equation}
a_1 = \sqrt{ \az^2 + 4 \laz}
,\quad
k_1 
=
\log \left[ \left(
1 + \frac{4\laz}{\az^2}
\right)^{-\frac12}
\right]
,\quad
g_1
=
\frac{\az}{a_1^2}
=
\frac{\az}{\az^2 + 4 \laz} .
\end{equation}
The thus obtained approximation to $P_0(h)$ is
\begin{equation}
\E^{k_1 + g_1 |h|^2 - \frac{\laz}{a_1^4} |h|^4}
=
\left(
1 + \frac{4\laz}{\az^2}
\right)^{-\frac12}\;
\E^{\frac{\az}{\az^2 + 4 \laz}|h|^2 
- \frac{\laz}{(\az^2 + 4 \laz)^2} |h|^4}
\end{equation}
By construction, this approximation is, for small $\laz$, 
exact to first order in $\laz$. 
But even at very large $\laz$
\begin{equation}
\frac{P_0(0)}{\E^{k(1)}} = \sqrt{\pi} + O (\laz^{-1/2}) .
\end{equation}
That is, the behaviour as a function of $\laz$ comes out correctly
as $\sim \laz^{-1/2}$, only the prefactor is too small. 
Note that $P_0(0)$ vanishes for large $\laz$, so the absolute 
error $|P_0(0) - \E^{k(1)}| = O (\laz^{-1/2})$ for large $\laz$. 
The difference of the true second moment to the one calculated
in this truncation is much bigger:
\begin{equation}
w_1 - g_1= \frac{1}{\sqrt{\pi\laz}} - \frac{\az}{\az^2 + 4 \laz} .
\end{equation}
but the qualitative behaviour, namely the vanishing at large $\laz$, 
is still reproduced correctly. The fourth order moment 
has the right scaling as a function of $\laz$, but the coefficient
is too large. 

The usual Polchinski hierarchy would not have given even a qualitatively
correct behaviour of these coefficients. Clearly, the adjustment of 
the propagator is crucial for large $\laz$ because the end result 
$w_1$ for the quadratic term is very different from the starting value $\az^{-1}$
at large $\laz$. 

However, the adjustment of the propagator alone is not sufficient for 
obtaining good asymptotics at large $\laz$ because 
truncations at higher $m$  improve the result only for small $\laz$, 
but not at large $\laz$.  
To see this, let us now truncate by setting $v_4=0$, so that \Ref{dickesei} becomes
\begin{equation}\label{v4trunc}
\dot v_3 = 4 a^{-1} \dot \chi v_2^2 .
\end{equation}
This corresponds to the usual truncation of the Polchinski equation,
where the six--point function is replaced by the tree contribution 
(see Figure \ref{fig3}).
used to obtain the one--loop equation for the four--point function. 
In the general case, \Ref{v4trunc} is converted into an integral 
equation which is then substituted back into \Ref{gelbesei}, 
leading to a flow equation for $v_2$ that is nonlocal in the scale parameter $s$. 
In the toy model, \Ref{gelbesei} and \Ref{v4trunc} can be combined
to 
\begin{equation}
v_2^2 \dot v_2
=
- \frac{9}{4}
v_3 \dot v_3 . 
\end{equation}
Because $v_3 <0$ at small $s>0$ by \Ref{dickesei},  
\begin{equation}
v_3 = -
\sqrt{\frac{8}{27} (\laz^3 - v_2^3)} 
\end{equation}
With this, \Ref{gelbesei} becomes 
\begin{equation}\label{bloedesei}
\dot v_2 = 9 a^{-1}\dot\chi \sqrt{\frac{8}{27} (\laz^3 - v_2^3)} ,
\end{equation}
and \Ref{bloedesei} and \Ref{aconvenient} now become a closed system.
Now there is a serious problem: $v_2$ decreases in the flow because 
$\dot\chi$ is negative, but this decrease never stops: even if $v_2 =0$, 
the $\laz$ term in \Ref{bloedesei} keeps $\dot v_2 $ negative, and 
the flow then goes to negative $v_2$ and becomes unstable. 
A numerical analysis of higher truncations shows a partial stabilization, but 
there is still an instability already at moderately large $\laz$. 
The reason for this behaviour will be discussed
in the next subsections.

\subsection{Discussion} 
\begin{figure}
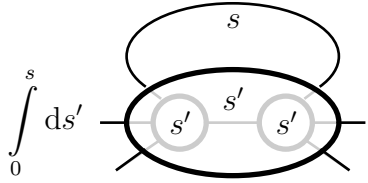

\begin{center}
\figplace{fig3}{0.0 in}{0.0 in}
\end{center}
\caption{The six--point tree diagram gets integrated over scales $s'<s$
and  contracted with a propagator at scale $s$, to give the right hand 
side of the flow equation for the four--point function at scale $s$.} 
\label{fig3}
\end{figure}

The above findings for the toy model can be understood easily 
in more general terms. The lowest truncation of the hierarchy is 
well--behaved in the toy model because the  
correction to $\az$ has a good sign, i.e.\ $a(s) $ remains 
positive for all $s$ because the $\la$--dependent terms 
get added to it (in addition, it is reasonably accurate).
The higher truncations become unstable
at large couplings because the hierarchy is effectively 
nonlocal in the flow parameter: if one truncates by putting 
the eight--point function to zero, the right hand side of the 
flow equation for the six--point 
function $v_3$ becomes a  tree diagram of four--point functions, 
so that the equation can be integrated, and the equation 
for the four--point function $v_2$ becomes (assuming $v_3 (0) =0$)
\begin{equation}
\frac{\del}{\del s} v_2 (s,\Psi)
=
\Delta_{D(s)}
\int_0 ^s \dd s'  \; 
v_2 (s',\Psi) \; F(s')\; v_2 (s',\Psi) 
\end{equation}
The graphical representation of this procedure is shown in 
Figure \ref{fig3}.
Because the entire flow history gets inserted on the right hand side, 
it does not help if $v_2$ decreases in the flow because there is 
always the contribution from small $s'$ where $v_2(s')$ is still 
close to $\laz$, hence large. This is the generalization of the 
right hand side of \Ref{bloedesei}, in which the $\laz$ term 
kept driving the flow to negative $v_2$ even at $v_2 =0$. 
Thus the nonlocality in the flow parameter makes it impossible
to see in this truncation that the effective coupling constant 
decreases, {\em but remains positive}, even at large initial coupling. 
In contrast, the 1PI equation is local in the flow parameter, 
and the just described ``infrared asymptotic freedom'' of the toy model at large $\laz$
is seen easily. Thus in conclusion,  to use the truncated equations at strong coupling
it is necessary that the scheme is stable
in the sense that asymptotic freedom is not violated in these truncations even
at large couplings.  

\subsection{The Wick ordered flow in the toy model}
In the light of the above discussion, one may ask
how an adaptive Wick ordered scheme would fare in this example, 
because the Wick ordered equation also has the property that it 
is local in the flow parameter.  As shown below, 
the Wick ordered flow turns out to be well--behaved. 
In fact,  already the initial Wick ordering provides a reasonable
approximation to the true result at large $\laz$. 

As already discussed in previous sections, 
Wick ordering has to be done self--consistently if one wants 
to keep the interaction free of quadratic terms. 
The initial Wick ordering provides, when done self--consistently, 
the solution of the mean--field theory (Hartree--Fock equations)
as a starting point for the flow. In the toy model,
self--consistent Wick ordering means finding $\aw$ such that
\begin{equation}\label{eq:169}
\az |\vphi|^2 + v(|\vphi|)
=
\aw |\vphi|^2 +  : v(|\vphi|):_{\aw^{-1}} .
\end{equation}
Because $v(|\vphi|) = \laz |\vphi|^4$
and $:v:_{\aw^{-1}} = \E^{- \aw^{-1} \del \delb} v$,
the analogue of \Ref{eq:1erlei} is  
\begin{equation}
|\vphi|^4
=
%\E^{- \aw^{-1} \del \delb} 
:|\vphi|^4 :_{\aw^{-1}}
+
4 \aw^{-1} |\vphi|^2
- 2 \aw^{-2}
\end{equation}
Comparison of the quadratic terms  in \Ref{eq:169} gives the self--consistency condition
\begin{equation}\label{toymf}
\az 
=
\aw -  4 \aw^{-1} \; \laz .
\end{equation}
The integral becomes
\begin{equation}
P_0(h) 
=
\sfrac{\az}{2\pi \I} \;
\E^{2 \laz \aw^{-2}}
\int_{\bC} \dd \bar\vphi \wedge \dd \vphi \;
\E^{-\aw |\vphi|^2 - :\laz |\vphi|^4 :_{\aw^{-1}} + h \bar \vphi + \bar h \vphi}
\end{equation}
and the standard shift operations then give
\begin{equation}\label{eq:covo}
P_0(h)
=
\sfrac{\az}{\aw}\;
\E^{2 \frac{\laz}{\aw^2}}\;
\E^{\frac{|h|^2}{\aw}}\;
\int \dd\mu_{\aw^{-1}} (\vphi) \; 
\E^{-\laz : |\vphi + \frac{h}{\aw}|^4:_{\aw^{-1}}}
\end{equation}
In contrast to mean--field equations in $d \ge 1$ dimensions, 
no integrations appear in the self--consistency relation \Ref{toymf}. 
It is simply a quadratic equation with solution
\renewcommand{\aw}{\tilde a}
\begin{equation}\label{eq:aw}
\aw 
=
\frac{\az}{2}
\left(
1 + \sqrt{1 + \frac{16 \laz}{\az^2}}
\right) .
\end{equation}
For small $\laz$, \Ref{eq:aw} gives an $O(\laz)$ correction to $\az$, 
and $\E^{2 \laz \aw^{-2}}$ gives the lowest order correction to $P_0 (0)$.
For large $\laz$, 
\begin{equation}\label{eq:lala}
\aw 
=
2 \sqrt{\laz} 
+
\frac{\az}{2}
+ 
\frac{\az^2}{16 \sqrt{\laz}}
+
O  (\laz^{-1}) .
\end{equation}
Originally, the strength of the coupling is given by the ``dimensionless'' ratio
$\laz/\az^2$. After the initial Wick ordering, this has decreased to 
\begin{equation}
\frac{\laz}{\aw^2} = \frac{1}{4} - O(\laz^{-1})
\end{equation}
Thus the ``strong coupling problem'' in this integral is removed 
already by the initial Wick ordering. 
Motivated by this, one can evaluate the remaining convolution integral
in \Ref{eq:covo} to first order in $\laz$, to get the approximation 
\begin{equation}
P_0 (h) 
\approx
\sfrac{\az}{\aw} \;
\E^{2 \laz /\aw^{2}} \; \E^{|h|^2 /\aw - \frac{\laz}{\aw^4} \, |h|^4}
\end{equation}
At large $\laz$, inserting \Ref{eq:lala} gives
\begin{equation}
P_0(0) \approx \frac{\sqrt{\E}}{2}\; \frac{\az}{\sqrt{\laz}} , \quad
w_2 \approx \half  \; \laz^{-1/2}\quad \mbox{ and }\quad
w_4 \approx - \frac{1}{16} \; \laz^{-1}
\end{equation}
Comparison with \Ref{eq:134} and \Ref{eq:136} shows that this approximation for 
$\sqrt{\laz} \; P_0(0)$ differs from the exact value by 8 \%,  the one for 
the prefactor of $\laz^{-1/2}$ in $w_1$  differs from the exact value by 12 \% , and 
the approximation for the prefactor of $\laz^{-1}$ in $w_2$ 
differs from the true value by less than 50 \% .
Note that these are the prefactors; the asymptotic behaviour as a function 
of $\laz $ comes out correctly. 
Thus a simple initial Wick ordering, together with first order perturbation theory, 
gives a  reasonable approximation for $P_0 (h)$ even at arbitrarily large $\laz$. 

Let us now consider the simplest truncation of the Wick ordered flow
with this initial condition.
For the toy example, the choice $r=d = a^{-1} \chi$ is possible,
so we use this to fix the Wick ordering covariance. It has the 
advantage that $r$ is determined without any self--consistency 
argument, but on the other hand, some tadpole terms remain in the equation.  
Setting $\chi (s) = 1-s$ with $0 \le s \le 1$ and 
expanding $w (\vphi) = \sum_{m \ge 2} w_m |\vphi|^{2m}$,
the RG equations for the Wick ordered functions read
\begin{eqnarray}
\dot a (s)
&=&
- (1-s) \; q(s) 
\label{wdota}\\
\dot w_2 (s)
&=&
- 20  (1-s) a (s)^{-2} \; w_2 (s)^2
\label{wdotw}\\
\dot k  (s)
&=&
- 8 (1-s)^3 \; a (s)^{-4} \; w_2 (s)^2
\\
\dot g (s)
&=&
a (s)^{-2} q (s)
\end{eqnarray}
The initial condition is given by the result of initial Wick ordering as
(fixing $a_0 =1$ in the definition of $P$ and recalling \Ref{eq:aw})
\begin{eqnarray}
a(0) 
&=& 
\aw = \half \left( 1 + \sqrt{1 + 16 \laz}\right)
\nonumber\\
w_2(0)
&=& 
\laz,
\quad 
k(0)
=
- \ln \aw + \frac{2 \laz}{\aw^2}, \quad
g(0)
=
{\aw}^{-1} .
\nonumber
\end{eqnarray}
and the function $q(s)$ is given by 
\begin{equation}
q(s) 
=
\frac{24 w_2 (s)^2\;a (s)^{-3} (1-s)^2}{1+ 4 a (s)^{-2} (1-s)^2 w_2 (s)}
\end{equation}
The numerator is the value -- in this toy model -- of the two--loop 
sunset diagram which is the only one contributing to $q(s)$ in 
the truncation $w_3 =0$. 
The denominator in the definition of $q(s)$ comes from a tadpole term, 
which is there because $\dot d - f = - a^{-2} \dot a \chi$. 
This is a special case of  the term $(1 - \Tad)^{-1}$ discussed in the previous
section. Again, the sign is nice in the toy example ---
 for $w_2 > 0$, the denominator is always at least $1$,
so this inversion does not introduce singularities. 
Only the equations \Ref{wdota} and \Ref{wdotw} are coupled; 
once their solution has been obtained, the other two simply follow by integration. 

%Equations  \Ref{wdota} and \Ref{wdotw} can be decoupled by 
Transform to the ``dimensionless'' functions
\begin{eqnarray}
\om (s) 
=
\frac{w_2(s)}{a(s)^2}
\quad
\mbox{ and }
\quad
\ell (s) = \ln \frac{a(s)}{a(0)}.
\end{eqnarray}
Then 
\begin{equation}\label{sehrschoen}
\dot \om (s)
=
-  \om(s)^2 \; \; 20 (1-s) \;
\Phi [(1-s)^2 \om(s) ] 
%s,\om(s))
%\frac{1 + \frac{8}{5}(1-s)^2 \om(s)}{1 + 4 (1-s)^2 \om(s)}
\end{equation}
with 
$
\Phi (x)
=
\frac{1+ \frac{8}{5} x}{1+4x}
$
and 
$\om (0)
=
\frac{\laz}{\tilde a^2} 
=
\frac{1}{4} - O(\laz^{-1})
$.
Given the solution of \Ref{sehrschoen}, $a(s) = a(0) \E^{\ell (s)}$ is obtained from 
the solution of
%{if there were a factor $2$ on the RHS of the eq. for $\dot \ell$,
%the result would differ from the exact one by less than .1 \%}
\begin{equation}
\dot \ell  (s)
=
- 24 \om(s)^2 \;
\frac{(1-s)^3}{1+4(1-s)^2 \om(s)} .
\end{equation}
Equation \Ref{sehrschoen} can be integrated explicitly 
but it is more instructive simply to note that $\om(s)$ is decreasing 
and $\om(s) > 0$ for all $s $ because $\om(0)> 0$ and
the right hand side of \Ref{sehrschoen} is negative 
and  vanishes for $\om(s) \to 0$ (the last property failed for the 
Polchinski--type equation). Thus $\Phi (x) $ appears only with $x \in [0, \om(0)]$. 
With $\om(0) = 1/4$ this implies $0.7 \le \Phi (\cdot ) \le 1$, hence
\begin{equation}
14 (1-s) \le - \frac{\dot \om (s)}{\om(s)^2} \le 20 (1-s) .
\end{equation}
Thus the solution for $\om(s)$ is bounded above and below by that of 
the prototypical equation for ``asymptotic freedom'' . 
Solving this equation and setting $s=1$ gives 
$\frac{4}{41} \le \om(1) \le \frac{4}{29} $.
Thus there is no instability in the Wick ordered scheme. 

To obtain the coefficient of $\laz^{-1}$ in $w_2$, 
the correct starting values are $\om(0) = 1/4$ and $a(0) = 2 \laz^{1/2}$. 
With this, the truncated equations give the approximation
\begin{equation}
w_2 \approx \frac{\om(1)}{a(0)^2\; \E^{2 \ell (1)}}
=
\frac{\om(1)}{4\; \E^{2 \ell (1)}} \; \laz^{-1}
=
0.033 \; %1/30
\laz^{-1}
\end{equation}
which differs from the exact value given in \Ref{eq:136} by about 20 \%. 
The solution for $g$ gives $w_1 \approx g(1) = 0.578 \laz^{-1/2}$. 
Comparison with \Ref{eq:136} shows a deviation of the coefficient
of $\laz^{-1/2}$ of about $2.5 \% $. 

In summary, the bad behaviour
of the Polchinski--type scheme at large
couplings is absent from the Wick ordered adaptive flow. 
The standard truncation $w_3=0$ is stable and reproduces the 
coefficients of the expansion of the exact result in inverse powers of $\laz$
well, i.e.\ with errors of order 1 -- 10 \% . 
Of course, the equation can also be used for all (not only large) values of $\laz$, 
where it has to be evaluated numerically. The stability properties are independent
of $\laz$. The numerical values are in good agreement with the exact result. 
A similarly good strong coupling behaviour of the Wick ordered functions
may also be expected in the more general examples discussed in \cite{Meden}. 

\subsection{Fermionic RG flows at strong coupling: general case}
\label{SCgen}
In this section I briefly discuss the question of well--definedness of 
the RG flow %hierarchy (and its solution) 
in fermionic systems at large couplings. 
It has been stated in many places that the RG hierarchy is a diagrammatic 
method that cannot be taken as a starting point at strong coupling. 
The true situation is, however, not really so bad: the very strategy of the RG 
of integrating out degrees of freedom in small steps makes the RG flow
well--defined also at large values of the initial coupling function. 
This is not a special feature of the differential equations but holds also 
for discrete RG flows, provided the stepsize is chosen small enough. 

To see this, consider a fermionic theory, as defined in the Introduction. 
By integrating the RGDE over short intervals, or using a discrete transformation, 
one can show that the generating function
\begin{equation}
\cW (C,\cV) (\phi) =  \log \int \dd\mu_C (\vphi) \; \E^{\la \cV(\vphi+\phi)}
= \log ( \mu_C * \E^{-\cV} ) (\phi)
\end{equation}
has the representation
\begin{equation}\label{eq:taeteraee}
\cW (C,\cV) 
=
\mu_C * \cV  + \sum_{p=2}^\infty \la^p \cW_p (C, \cV, \ldots \cV)
\end{equation}
where in $\cW_p$, the argument $\cV$ appears $p$ times and 
\begin{equation}
\cW_p(C,\cV_1, \ldots, \cV_p) (\Psi)
=
\frac{1}{p!}
\sum_{T \in \cT_p}
\int \dd P (\si)
\Leval
\E^{\Delta_C [M(\si)]} 
\pli_{t \in T} 
\Delta_C^{(t)}
\pli_{q=1}^p
\cV_q (\Psi_q)
\Reval
\end{equation}
where $\cT_p$ is the set of all trees on $p$ vertices, and $P$ is a probability 
measure (i.e.\ positive and normalized, $\int \dd P(\si ) = 1$), 
and $M(\si)$ is a positive definite $p \times p$ matrix with diagonal 
entries equal to $1$. 
Both $P$ and $M$ can be written down explicitly, but only the 
above--mentioned properties matter (see \cite{SalWiecz},
where the above formula is derived and explained in more detail).
Eq.\ \Ref{eq:taeteraee} can be thought of as a resummation of the perturbation 
expansion in terms of trees. Because $\frac{1}{p!}|\cT_p| = \frac{1}{p!}p^{p-2}\le \E^p$,
the sum converges if the contribution of every tree $T$ can be bounded by $\const^p$. 
For fermions, this can be shown using determinant bounds (see, e.g.\ \cite{FKT,SalWiecz,PeSaUV}).
The crucial point for the present discussion is that for a short--range interaction $\cV$
sufficient conditions for convergence are that  the determinant constant $\de_C$ of $C$
(see \cite{PeSaUV})
and the decay constant $\al_C = \sup_X \int |C(X,Y)| \dd Y$ are so small that 
$|\la| \al_C \ll 1$ . 
In a RG flow with stepsize $\veps$, the covariance $C$ is split into many small parts $C_\veps$
and in every RG step, the $C$ in the above formula is replaced by $C_\veps$. 
Because $\al_{C_\veps} \sim \veps$ and $\de_{C_\veps} \sim \veps^{1/2}$, 
the above convergence condition can be satisfied for any $\la$ by making the 
RG stepsize $\veps $ small enough. 
Thus, for small enough $\veps$
the first steps in the RG integration are given by convergent perturbation theory 
in $\la $, and therefore the RG flow can be started at large coupling. The really 
interesting analytical and physical question is whether
a large, repulsive local initial interaction indeed  decreases in the
course of the flow (i.e.\ it gets screened),
and whether the rate of this decrease is so fast that the flow
can be controlled in the regime where the local coupling is 
not yet small, as is the case in the toy model. 

One may indeed expect that a strong on--site repulsion gets 
screened in the course of an RG flow, because even a hard--core condition
gets softened, thus effectively weaker when one considers 
larger blocks on the lattice. Namely, if there is a strong on--site repulsion, 
double occupancy of single sites is exponentially suppressed, 
but hopping into a block of $m^d$ lattice sites, $m > 1$, is still possible
because of local fluctuations in the number density, even if the 
overall density is fixed. 
A standard example for a stochastic system with a hopping term and
a hard--core repulsion is the asymmetric exclusion process. It was shown in 
\cite{LQSY} that averaging in a suitable way effectively allows to 
{\em remove} the hard--core constraint in this model and still 
get precise asymptotics of the decay of correlations. 
Because the truncations of the correlation hierarchy in this example
have the property that they provide successive upper and lower bounds
for the exact solution, one can obtain their asymptotic properties
in some detail \cite{LQSY}. For fermion systems, such monotonicity
properties are in general not known, 
nevertheless averaging, such as provided by the RG, 
seems a very promising  strategy to see screening effects.
In fact, a low--order evaluation of the flow at very large scales 
in the two--dimensional Hubbard model shows the screening,
as well as the generation of the usual antiferromagnetic interaction term
\cite{MSunpub}.

The above argument is about the issue of 
the behaviour of an {\em initial} strong short--range interaction.  
The ``flow to strong coupling'' at low scales that has been observed in 
studies of initially weakly coupled systems is an independent phenomenon
(as discussed in \cite{SH}, such flows may be kept under control for a while
by phase space arguments, but eventually, the growth of the coupling function
wins). The growth of certain parts of the coupling function
in these flows is really due to the emergence of order parameter fluctuations
that become more and more long--range, hence lead to fermionic interaction 
vertices that develop singularities in momentum space at a scale 
where symmetries get broken. In situations where, e.g.\ 
the singular part of the fermionic four--point  function 
can be parametrized by an exchange of  a boson that becomes 
massless at a certain scale, approximations that set the momentum 
of this boson to zero give a coupling constant of order $m_{\rm Boson}^{-2}$
which ``runs to strong coupling'' and diverges at a certain scale. 
Whether the boson can really become massless depends on the situation,
in particular the dimensionality of the system. Moreover, in systems with 
long--range interactions, the Anderson--Higgs mechanism may interfere. 
However, unless this is the case, the Goldstone theorem implies such 
singularities whenever continuous symmetries are broken.  
Because such singularities occur only at points, their effect is very 
different from that of a strong local interaction, which is equally large
everywhere in momentum space.  

\section{Conclusions}
The RG equations developed here allow for a dynamical adjustment of propagators 
in a convenient and flexible way. In particular, they allow
for countinuously adjusted scale decompositions, which are useful in Fermi surface studies,
when one wants to avoid the use of counterterms, or 
fix the density (instead of the 
chemical potential) in RG flows.
A numerical study of the adaptive flow equation discussed here, with 
fixed density,  was done for the two--dimensional 
Hubbard model in the appendix of \cite{HSFR}. A mathematical study 
for the case of regular Fermi surfaces is in \cite{PeSa}. 
Taking the Fermi surface deformation into account is important
also for the case of singular Fermi surfaces, where the question of 
pinning of the Fermi surface at Van Hove singularities can be addressed
using such flows. This situation is studied in \cite{SFS} using counterterms. 
The adaptive method is expected to lead to useful results
also here. 

In flows with fermion and boson fields, a generalization of the 
strategy for the adaptive flow becomes obvious: quartic terms in the 
fermions contribute to the boson propagators, so the dynamical 
adjustment of the boson propagator allows to take out also 
certain terms from the fermionic four--point function. 
A related method has already been used in the 1PI scheme in \cite{Baier}; 
a straightforward generalization of the method developed here
will allow to do it in any scheme. 

Finally, the role of the propagator adjustment for the strong coupling 
behaviour of flows was investigated for a very simple toy model. 
% in the ``unreasonable effectiveness''
%of the 1PI RG scheme at strong coupling, in a very simple toy model, 
%was investigated. 
It was shown that the Wick ordered scheme with 
dynamic propagator adjustment correctly reproduces the asymptotic 
behaviour for arbitrarily large couplings,  and already its simplest truncation
gives the coefficients of the leading term in the asymptotic series
with an error of order 1 -- 10 \%  (depending on the
order of the moment).
The analysis made clear that the adjustment of propagators is indeed
crucial for the success of a scheme at strong coupling, but that in addition, 
a stability property  of the four--point flow is necessary. 
Namely, the RG equations must be such that asymptotic freedom 
holds for the solution also if the flow is started at a large value of the coupling. 
The Polchinski scheme does not have the second property, even with an adapted
propagator, but the 1PI and Wick ordered schemes do. 

Needless to say, the toy example is very special because there, 
the signs work out  such that a large initial coupling $\laz$ decreases to a very small
final value, and the quadratic term $\az$  increases to a large final value. 
Because only the combination $\la/a^2$ matters, this effectively leads to 
a very rapid transition to a small--coupling situation in the flow, provided
the above--mentioned stability holds. This explains the success of the 
1PI and Wick ordered scheme. In the latter, already the initial Wick ordering 
removes the strong coupling problem right away, if it is done self-consistently. 
It would be too much to hope for such benign signs in more general (and more
interesting) models,  but, as discussed in some detail in Section \ref{SCgen}, 
the flow equations are well--defined at strong coupling (contrary to what  
has often been stated in the literature) 
and there  are reasons to be optimistic about screening
of initially strong local repulsions in RG flows.

\bigskip\noindent
{\bf Remarks and Acknowledgements. } 
My research on this topic started in 1998, but
this paper has been held back for a long time for various reasons.
%the most important one being inappropriate perfectionism. 
I would like to thank Carsten Honerkamp and Walter Metzner for discussions, 
and the KITP, Santa Barbara, especially the program
{\em Realistic Theories of correlated electron systems} 
and the Erwin Schr\" odinger Institut, Vienna, 
for hospitality and financial support
in various stages of the work. During the last year, it was also supported by
DFG grant Sa 1362/1--1 and NSERC of Canada.

\end{document}